\documentclass[journal]{IEEEtran}  % Option "lettersize" causes an "Unused global option(s)" warning.
\IEEEoverridecommandlockouts
\usepackage{amsmath,amsfonts}
\usepackage{bm}
\usepackage{amssymb}
\usepackage{hhline}
\usepackage{algorithmic}
\usepackage{algorithm}
\usepackage{array}
\usepackage[dvipsnames]{xcolor}
\usepackage{hyperref}
\ifCLASSOPTIONcompsoc
    \usepackage[caption=false, font=normalsize, labelfont=sf, textfont=sf]{subfig}
\else
\usepackage[caption=false, font=footnotesize]{subfig}
\fi
\usepackage{textcomp}
\usepackage{stfloats}
\usepackage{url}
\usepackage{verbatim}
\usepackage{graphicx}
\usepackage{cite}
\def\etal{\textit{et al.}}

\newcolumntype{P}[1]{>{\centering\arraybackslash}p{#1}}

\begin{document}

\title{Subjective Quality Assessment of Compressed Tone-Mapped High Dynamic Range Videos}

\author{Abhinau K. Venkataramanan and Alan C. Bovik,~\IEEEmembership{Life Fellow,~IEEE}
        % <-this % stops a space
\thanks{This research was sponsored by a grant number 2019844 from the National Science Foundation AI Institute for Foundations of Machine Learning (IFML).}}% <-this % stops a space

\markboth{Journal of \LaTeX\ Class Files,~Vol.~XX, No.~X, Month~20XX}{}


\maketitle

\begin{abstract}
High Dynamic Range (HDR) videos are able to represent wider ranges of contrasts and colors than Standard Dynamic Range (SDR) videos, giving more vivid experiences. Due to this, HDR videos are expected to grow into the dominant video modality of the future. However, HDR videos are incompatible with existing SDR displays, which form the majority of affordable consumer displays on the market. Because of this, HDR videos must be processed by tone-mapping them to reduced bit-depths to service a broad swath of SDR-limited video consumers. Here, we analyze the impact of tone-mapping operators on the visual quality of streaming HDR videos. To this end, we built the first large-scale subjectively annotated open-source database of compressed tone-mapped HDR videos, containing 15,000 tone-mapped sequences derived from 40 unique HDR source contents. The videos in the database were labeled with more than 750,000 subjective quality annotations, collected from more than 1,600 unique human observers. We demonstrate the usefulness of the new subjective database by benchmarking objective models of visual quality on it. We envision that the new LIVE Tone-Mapped HDR (LIVE-TMHDR) database will enable significant progress on HDR video tone mapping and quality assessment in the future. To this end, we make the database freely available to the community at \url{https://live.ece.utexas.edu/research/LIVE_TMHDR/index.html}.
\end{abstract}


\section{Introduction}
\label{sec:introduction}
The real world presents the human visual system (HVS) with a wide range of luminance (brightness) ranges, even in ordinary settings. For example, the luminance of starlight is a mere 0.0003~cd/\(\text{m}^2\) (nits), while the luminance of bright sunlight on a clear day can reach as high as 30,000 nits. Due to the iris' control on the pupil size and other adaptive gain control mechanisms, the human eye is able to perceive wide ranges of brightnesses, from around \(10^{-6}\) nits to \(10^8\) nits. However, legacy imaging and display systems are only capable of capturing or generating narrower ranges of brightnesses, up to the order of 100 nits. Such systems are referred to as low or standard dynamic range systems (SDR). Another limitation of SDR systems is that they span only around 35\% of all visible gamut of colors. Examples of legacy SDR standards include ITU BT. 709 \cite{ref:rec_709} and sRGB \cite{ref:srgb}.

To expand the scope of imaging and display systems to meet the capabilities of human vision, high dynamic range (HDR) imagers and displays have been developed over the years. Modern HDR standards such as ITU BT. 2100 \cite{ref:rec_2100} are able to represent luminances in the range of \(10^{-4}\) to \(10^4\) nits, and wider color gamuts (WCGs) that can represent about 75\% of the volume of visible colors. This is achieved by combining two or more images captured at different exposure settings using computational imaging techniques. To effectively encode and transmit wide ranges of brightnesses, the captured image signals are modified by ``opto-electrical transfer functions'' (OETFs), which generalize the notion of gamma correction of legacy Cathode Ray Tube (CRT) displays.

Two OETFs have been included in the BT. 2100 standard for this purpose: the Perceptual Quantizer (PQ) \cite{ref:pq} and Hybrid Log-Gamma (HLG) \cite{ref:hlg}. PQ is a ``forward-compatible'' function capable of encoding luminances up to \(10^4\) nits and is typically used by professional studios that deliver high-quality HDR content. The PQ encoding function is part of the HDR10 \cite{ref:hdr10} and HDR10+ \cite{ref:hdr10_plus} standards. Conversely, HLG is designed to be ``backward-compatible'' with SDR standards, by including ``gamma'' curves similar to those used in SDR. While the HLG standard does not define a peak luminance, a nominal value of 1000 nits is commonly be used. HLG has found adoption in satellite television networks to enable HDR content delivery \cite{ref:hlg_satellite}. Both PQ and HLG are supported by the emerging Dolby Vision standard \cite{ref:dolbyvision}.

A key reason why the widespread streaming of HDR video content is limited is the scarcity of deployed true HDR displays. The BT.2100 standard defines true HDR systems as those that support at least 1000 nits \cite{ref:rec_2100}. However, most affordable HDR displays do not meet this threshold, managing peak brightnesses of 800 nits or less \cite{ref:hdr_tv_survey}. Indeed, a large percentage of legacy displays in current use only support SDR formats \cite{ref:sdr_majority}. Therefore, to make HDR video content accessible to a wide range of consumers, it is necessary to ``down-convert'' them to the SDR range in a perceptually acceptable way. Such a process is called ``tone-mapping.''

Many tone-mapping methods have been proposed in the literature, and a brief review of them is presented in Section \ref{sec:database_tonemapping}. However, due to the limited dynamic range of SDR systems, even the latest tone mapping operations introduce visual distortions. These distortions often take the form of contrast losses or gains, reduction of visual details, especially in dark or bright regions, or introduced visual artifacts, which can be particularly visible at the extreme ends of the dynamic range \cite{ref:hdrmax_spl}. Moreover, due to the remapping of color across dynamic ranges and the WCGs used by HDR, chromatic distortions may also occur, particularly in the form of hue shifts and chroma-clipping artifacts \cite{ref:cadik} \cite{ref:gma_review}. Finally, the increased bit depths of HDR videos implies that streaming videos over the internet, which is already extremely bandwidth intensive, incurs high bandwidth costs, necessitates lossy video compression. This adds another layer of distortions in the form of blocking, banding, crushing of detail, amid numerous temporal distortions \cite{ref:compression_effects}.

Here, we study the deeply connected problems of subjective perception and objective prediction of the visual qualities of compressed tone-mapped videos. In particular, we created the LIVE-Tone Mapped High Dynamic Range Video (LIVE-TMHDR) Database, which is the first large-scale publicly available database of HDR videos that have been subjected to tone-mapping and compression. LIVE-TMHDR contains 40 source HDR videos, which have been tone-mapped using a diverse collection of tone-mapping operators followed by various degrees of H.264 compression, yielding 15,000 distorted videos. We also obtained a large number of reliable subjective annotations of visual quailty by conducting an online crowdsourced subjective study. We then processed the collected subjective quality annotations to obtain quality labels for each video, which used to we analyze the perceptual outcomes and relative performance of of the compared tone-mapping operators. Finally, we demonstrate the usefulness of the new database by evaluating a wide variety of state-of-the-art video quality models on it.

The rest of the paper is organized as follows. Section \ref{sec:background} provides background regarding the literature of subjective quality assessment of tone-mapped pictures and videos. Section \ref{sec:database} describes the construction of the LIVE-TMHDR database, while Section \ref{sec:study} describes the methodology of the crowdsourced subjective study. Section \ref{sec:data_analysis} describes the methods used to process and analyze the subjective data, and Section \ref{sec:evaluation} describes the outcomes of evaluating many state-of-the-art (SOTA) objective video quality models on the LIVE-TMHDR database. The paper concludes in Section \ref{sec:conclusion}.
\section{Background}
\label{sec:background}
Over the years, many efforts have been made to understand the impact of various tone-mapping operators (TMOs) on visual quality, often with the aim of comparing TMOs and identifying the ``best'' method. A common methodology used in these subjective experiments are pairwise comparison (PC) tests, which are considered suitable due because of the often subtle differences of the outcomes of TMOs. In studies that follow the PC protocol, subjects are tasked with choosing the better of two tone-mapped versions of a same visual content. In some cases, cross-content evaluations may be performed, allowing for the mapping of PC preferences to an absolute opinion scale.

Drago \etal \cite{ref:drago} conducted the first comparison of TMOs by applying 6 TMOs on 4 HDR source images. The results of PC tests were analyzed using the INDSCAL \cite{ref:indscal} model towards understanding the dependencies of subject preferences on detail preservation and visual naturalness. Similar PC tests were conducted by Kuang \etal \cite{ref:kuang} and Ledda \etal \cite{ref:ledda}. Kuang \etal applied eight TMOs on ten scenes and conducted PC tests using 33 and 23 subjects in two experiments. Thurstone's method \cite{ref:thurstone} was used to map PC results to absolute opinion scores. Among the TMOs compared, it was found that the algorithms developed by Durand \etal (``Durand02'') \cite{ref:durand02} and Reinhard \etal (``Reinhard02'') \cite{ref:reinhard02} achieved the highest human quality ratings. On the other hand, Ledda \etal \cite{ref:ledda} applied eight TMOs to 23 scenes. The results of the PC comparisons revealed that the iCAM \cite{ref:icam} and Reinhard02 methods were the most preferred.

Yoshida \etal \cite{ref:yoshida} applied seven TMOs on two static scenes, yielding 14 tone-mapped pictures. Detailed subjective ratings were obtained, describing the preservation of visual attributes, such as brightness, contrast, detail reproduction in bright and dark regions, and overall naturalness. The results showed that the choice of the ``best TMO'' was dependent on the visual attribute being rated. Cadik \etal \cite{ref:cadik} conducted a similar, yet more extensive, analysis using fourteen TMOs applied on three scenes. Subjective ratings were collected to describe the overall perceived visual quality of each picture along with attributes such as brightness, color, contrast, detail reproduction, and the presence of artifacts. An analysis of the quality ratings yielded the unexpected result that global TMOs that use simple constant tone curves generally outperformed local TMOs that utilize local processing. Krasula \etal \cite{ref:krasula_tm_db} compared the perceived quality of tone-mapped pictures of natural and synthetic contents, both in the presence and absence of reference pictures. The results of the human experiments showed that subjective preferences depend on the availability of HDR reference images for natural scenes, but not for synthetic scenes.

Finally, Eilertsen \etal \cite{ref:eilertsen_comp16} applied eleven TMOs to six source HDR videos, creating a database of 66 videos. PC tests were used to collection human comparisons of the videos, and detailed feedback was collected regarding the distortions introduced by each TMO. To the best of our knowledge, the database of tone-mapped videos and subjective ratings has not been made publicly available. An extensive review of 26 TMOs for HDR video was also performed by Eilertsen \etal \cite{ref:eilertsen_review17}, though a subjective study was not performed.

In addition to making TMO comparisons, a variety of subjective databases have been created to aid in the process of developing quality algorithms for tone-mapped pictures and videos. Yeganeh \etal \cite{ref:tmqi} developed a database of 120 pictures by applying eight TMOs on fifteen source HDR scenes. The database was rated using a ranking method and used to evaluate the performance of the Tone-Mapped Quality Index (TMQI) proposed in the same work. The ESPL-LIVE HDR database \cite{ref:espl_live} is a large-scale database of 747 tone-mapped pictures, consisting of 605 unique contents processed by four TMOs and five multi-exposure fusion (MEF) methods. ESPL-LIVE was used to validate the HIGRADE \cite{ref:higrade} objective picture quality model. Finally, RV-TMO \cite{ref:rvtmo} is a new large-scale database consisting of 1000 tone-mapped pictures generated by applying four TMOs to 250 source HDR images. The outcomes were rated by human subjects using a PC protocol. The RV-TMO database was used to evaluate a suite of competitive visual quality models.

A summary of the databases discussed in this section is presented in Table \ref{tab:prev_dbs}. By comparison, the new LIVE-TMHDR Database that we describe here is the first publicly available subjective database of tone-mapped HDR videos, and the largest in terms of the number of tone-mapped contents (including picture databases), embodying a total of 15,000 distorted videos. Moreover, LIVE-TMHDR is the first database to be explicitly focused on generating and studying distortions related to tone-mapping in the presence of video compression, rather than only comparing TMO outcomes on pristine contents.

\begin{table}[t]
    \centering
    \caption{Summary of Tone-Mapped Picture and Video Quality Databases in the Literature. Publicly Available Databases Are Bold-Faced.}
    \label{tab:prev_dbs}
    \begin{tabular}{|c|c|c|c|c|c|}
    \hline
    Name & Type & Year & Sources & TMOs & Distorted \\
    \hline
    Drago & Pictures & 2003 & 4 & 6 & 24 \\
    \textbf{Kuang} & Pictures & 2004 & 10 & 8 & 80\\
    \textbf{Ledda} & Pictures & 2005 & 23 & 6 & 138\\
    Yoshida & Pictures & 2005 & 2 & 7 & 14 \\
    \textbf{Cadik} & Pictures & 2008 & 3 & 14 & 42 \\
    \textbf{Yeganeh} & Pictures & 2013 & 15 & 8 & 120\\
    Eilertsen & Videos & 2016 & 6 & 11 & 66 \\
    \textbf{Krasula} & Pictures & 2017 & 20 & 5 & 180 \\
    \textbf{ESPL-LIVE} & Pictures & 2017 & 605 & 4 & 747\\
    \textbf{RV-TMO} & Pictures & 2022 & 250 & 4 & 1000\\
    \hline
    \end{tabular}
\end{table}

\section{The LIVE-TM-HDR Database}
\label{sec:database}

In this section, we present the new LIVE-TM-HDR database. The primary objective of the database is to provide a rich testing ground for understanding and modeling the interplay between distortions introduced by tone mapping and compression. Both sources of distortion are ubiquitous in the delivery of HDR content to SDR displays, which comprises the majority of visual consumption by consumers viewing HDR.

To summarize, with details to follow, we constructed the database by assembling a diverse set of HDR video source contents, which we then distorted by first tone-mapping them to convert them into SDR, then by lossily compressing them. We identified a set of ten prominent open-source TMOs, each of which we deployed to process each source video using four spatial parameter values and three ``temporal modes.'' We also applied two proprietary TMOs each configured in two temporal modes, and we commissioned a professional colorist to manually tone-map the HDR videos. Finally, since compression is a ubiquitous distortion of videos delivered over the internet, we applied three levels of lossy compression on all of the tone-mapped videos.

It is important to note that in the subjective experiment (Section \ref{sec:study}), none of the original, unprocessed, HDR source contents were included for subject annotation. The reason for this is two-fold. First, very few of the workers on the crowdsourcing platform that we used (Amazon Mechanical Turk, or AMT) have access to HDR displays. Even among those that do, it would be too disruptive to expect workers to switch displays (or display modes) to view both HDR and SDR videos during a single session. Second, the target use case is the streaming of HDR videos to consumers who do not have access to HDR displays, and who would never view true HDR contents.

Overall, we created a large-scale distorted video database containing 15,000 tone-mapped videos that include highly diverse spatiotemporal distortions caused both by tone-mapping and by compression. Details regarding the construction of the database are provided in the following subsections.

\subsection{Source Contents}
\label{sec:database_sources}

\begin{figure*}[ht]
    \subfloat[Sample PGC (PQ) source contents\label{fig:pq_sample_frames}]{%
      \includegraphics[width=\linewidth]{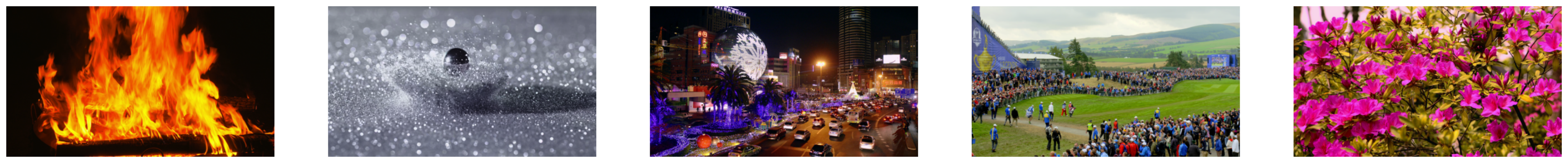}
    }
    \\
    \subfloat[Sample UGC (HLG) source contents\label{fig:hlg_sample_frames}]{%
      \includegraphics[width=\linewidth]{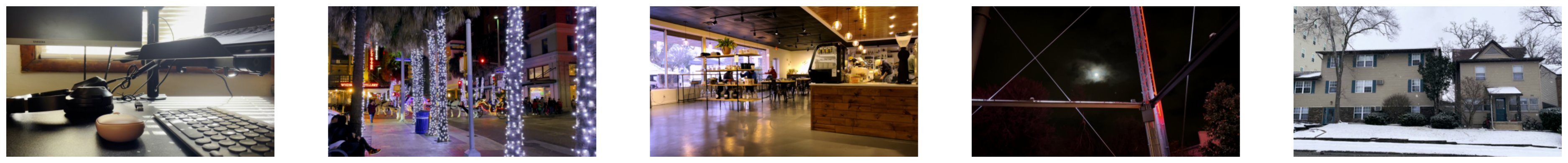}
    }
    \caption{Sample frames from ten HDR source contents, tone-mapped by a professional colorist.}
    \label{fig:sample_frames}
\end{figure*}

We gathered a diverse set of 40 HDR source contents as the basis of the LIVE-TMHDR Database. These videos include exemplars of both low frame rate (LFR) user-generated content (UGC) and high frame rate (HFR) professionally-generated content (PGC). 

The PGC sub-database consisted of 20 HDR videos sourced from the LIVE-HDR database \cite{ref:live_hdr}, and encoded using the Perceptual Quantizer (PQ) transfer function. These videos were in turn obtained from various repositories, such as the Consumer Digital Video Library (CDVL) \cite{ref:cdvl}, the SJTU-HDR database \cite{ref:sjtu_hdr}, and 4KMedia \cite{ref:4kmedia}. These PGC source videos are all 4K high frame rate (50 or 60 fps) HDR videos. 4K video streaming is estimated to require about 7 GB of data per hour, which is double the requirement of 1080p video streaming \cite{ref:sandvine2023}. Such high data requirements pose challenges during both real-world streaming scenarios, and online crowdsourced subjective experiments. Moreover, since 4K displays tend to be more expensive than HD displays, we expect fewer workers on crowdsourced platforms to have access to them. Therefore, we decided to restrict our database to 1080p videos, and downscaled all the source contents using Lanczos interpolation.

The UGC portion of the database was obtained by sourcing 20 HDR videos from amateur iPhone users, each of which was encoded using the Hybrid-Log Gamma (HLG) transfer function. Of these twenty videos, twelve were capturing using iPhone 12s, while eight were captured on iPhone 13s. Since the iPhone 13 supports the recording of HDR videos at 4K resolution and a frame rate of 60 fps, these videos were downsampled in space and time to a resolution of 1080p and a framerate of 30 fps, to match the videos filmed using iPhone 12s.

All of the source contents collected in this manner were cropped to a maximum duration of 10 seconds, yielding a database of 40 source contents consisting of both PQ and HLG videos, at framerates of 30, 50, and 60 fps. Screengrabs of sample frames from among the 40 HDR contents, all tone-mapped by a professional colorist, are shown in Fig. \ref{fig:sample_frames}. Further details regarding the manual tone-mapping by the professional colorist are provided in Section \ref{sec:database_tonemapping}. 

To characterize the set of source HDR contents, we computed three low-level descriptors - Spatial Information (SI), Temporal Information (TI), as defined in \cite{ref:vqeg_siti}, and Colorfulness (CF), as defined in \cite{ref:cf}. To compute these features from HDR videos, we follow the recommendations made by the Video Quality Experts Group (VQEG) \cite{ref:vqeg_siti} and convert all videos into PQ encodings, followed by rescaling the features to the range 0-255.

\begin{figure}[tb]
    \subfloat[SI vs TI\label{fig:si_v_ti}]{%
      \includegraphics[width=0.48\linewidth]{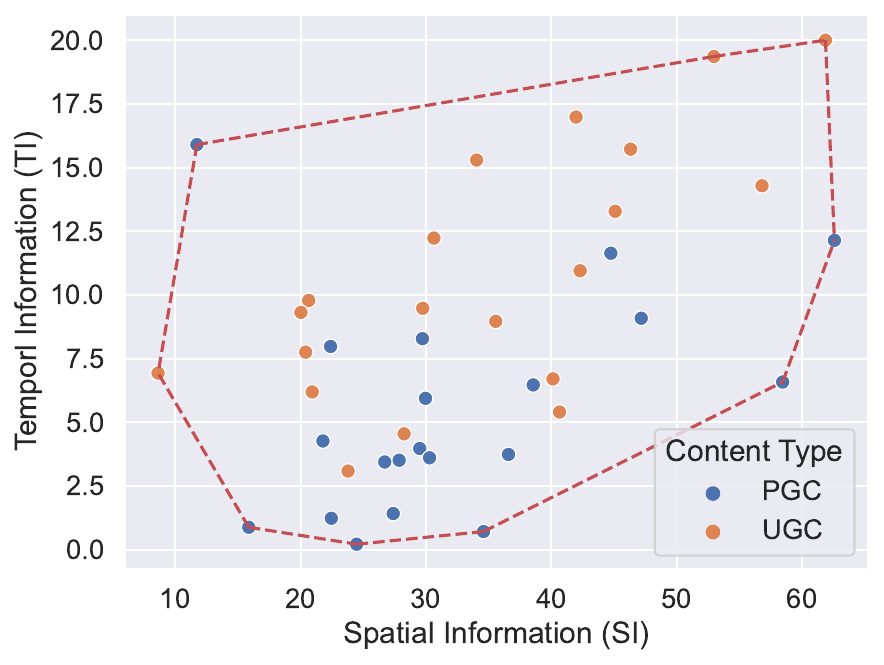}
    }
    \hfill
    \subfloat[SI vs CF\label{fig:si_v_cf}]{%
      \includegraphics[width=0.48\linewidth]{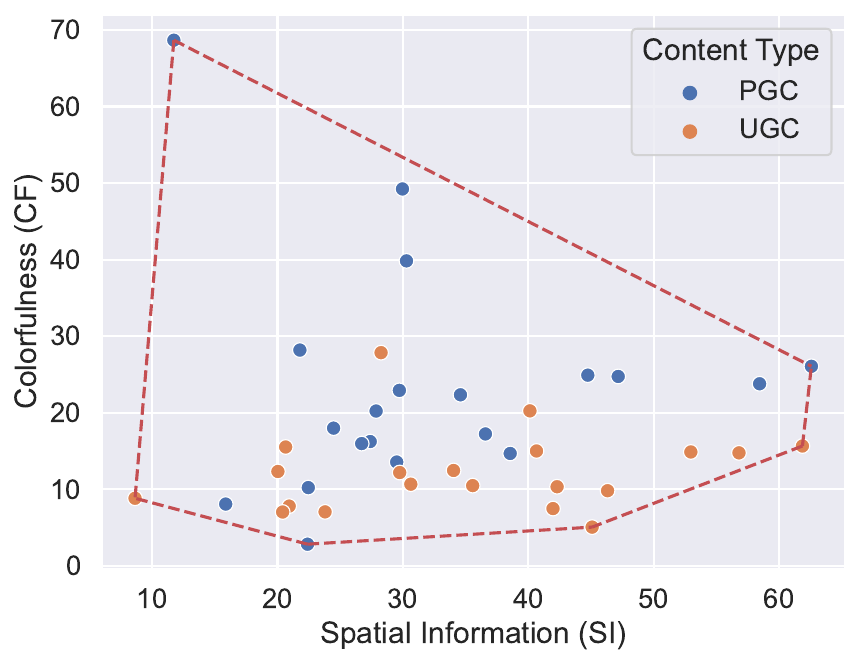}
    }
    \caption{Convex-hull analysis of low-level descriptors of HDR source videos.}
    \label{fig:si_ti_cf}
\end{figure}

As in \cite{ref:live_hdr} and \cite{ref:ugc_vqa}, we used these low-level descriptors to illustrate the content diversity of the source videos in the LIVE-TM-HDR database. Convex hull diagrams showing the distributions of SI, TI, and CF are shown in Fig. \ref{fig:si_ti_cf}, from which it may be observed that the source contents cover a broad range of spatial and temporal features.

Finally, as alluded to above, we commissioned a professional colorist to manually grade, i.e., tone-map, the HDR videos with the objective of best reproducing the appearance of the source contents. These expert tone-mapped videos provide a baseline against which the distortions introduced by various TMO algorithms may be evaluated.

\subsection{Tone-Mapping Operators}
\label{sec:database_tonemapping}
The primary distortions that we study here are those induced by TMOs and by compression, as well as combinations of these, which may be regarded as novel distortions. HDR videos must adhere to the ITU-BT.2100 standard. When the HDR videos are converted into SDR format, they adhere to the sRGB standard. We chose sRGB over newer WCG SDR standards such as BT.2020, since HD SDR video typically does not use WCGs.

Tone-mapping refers to the process of ``down-converting'' HDR videos to a reduced SDR range of luminances, with a particular focus on reproducing, and perhaps enhancing, local contrast, global contrast, and color appearance. Broadly, tone-mapping operators (TMOs) are classified into two types - global and local.

Global TMOs typically apply a constant compressive point non-linearity on entire frames. By comparison, local TMOs use adaptive compressive non-linearities that capture information regarding local contrast and luminance.

The ITU-BT.2100 standard, which all HDR sources in the new database adhere to, uses a wide color gamut (WCG) to represent color. Therefore, a gamut mapping algorithm (GMA) is required to restrict the colors in the SDR videos to the narrow BT.709 color gamut, which is used by the sRGB standard.  

Gamut mapping is a broad area of research that includes techniques such as gamut clipping \cite{ref:mindist_gma} \cite{ref:katoh_clipping} and gamut compression \cite{ref:morovic_gma} \cite{ref:sigmoid_gma}, which may be linear or non-linear. GMAs may also depend on the image, and may use measurements such as color difference (\(\Delta E\)), lightness mapping, and hue linearity. A thorough review of GMAs is provided in \cite{ref:gma_review}. Because we limit our focus to tone-mapping, we apply a simple linear gamut clipping method in the XYZ color space, as described in Section 2 of ITU Recommendation BT.2407 \cite{ref:itu_gma}.

So that we could test and compare a wide range of authentic tone-mapping operators and their characteristic distortions, we developed an open-source Python library of ten TMOs that deploy diverse ways of accomplishing tone-mapping. These TMOs include such diverse methods as pointwise non-linear transforms, multi-scale decompositions, clustering, and deep neural networks.

We will refer to each TMO by author and year of publication. Brief descriptions of each of the ten TMOs follows:

\begin{enumerate}

\item Hable \cite{ref:hable} - The Hable TMO, implemented as a ``filter'' in FFmpeg \cite{ref:ffmpeg}, is a parameter-free pointwise non-linear transform originally designed for use in the video game \textit{Uncharted 2}. The Hable filter has found wide use due to its inclusion in FFmpeg, making it a prime candidate for our experiments. 

\item Reinhard02 \cite{ref:reinhard02} - The global version of the Reinhard02 TMO, which we use here, applies a point non-linearity to map luminances from HDR to SDR. Chrominance values are scaled proportionally with the luminances.

\item Durand02 \cite{ref:durand02} - The Durand02 TMO uses a ``fast bilateral filter'' to decompose the luminances of HDR frames into ``base'' and ``detail'' layers, corresponding to illumination and reflectance. The base layer is linearly scaled in the logarithmic domain to achieve a predetermined contrast, then the details are reintroduced. Chrominance values are scaled proportionally with the luminances.

\item Shan12 \cite{ref:shan12} - The Shan12 TMO utilizes an edge-aware stationary wavelet transform (SWT) \cite{ref:swt} to decompose HDR frames. Appropriate gains are applied to the subbands, then the wavelet transform is reversed, yielding the SDR frames.

\item Reinhard12 \cite{ref:reinhard12} - The Reinhard12 TMO uses color-appearance models, applied in a local manner, to predict the cone responses in a human eye when viewing the HDR frame. The SDR frame is then generated to produce analogous cone responses in the SDR range, towards best reproducing the appearance of the HDR frame.

\item Eilertsen15 \cite{ref:eilertsen15} - The Eilertsen15 TMO is a multi-stage algorithm that first applies a ``fast detail extraction'' method to obtain a base-detail decomposition similar to that in \cite{ref:durand02}. Then, a contrast distortion objective is minimized to derive a dynamic tone-curve, which is then used to tone-map the input frame. A model of camera noise is used to adapt tone curves to reduce noise visibility.

\item Oskarsson17 \cite{ref:oskarsson17} - The Oskarsson17 TMO uses Dynamic Programming to cluster values in the input image channels. The mapping to clusters is done in the log domain, to form a tone curve that maps HDR values to the SDR range.

\item Rana19 \cite{ref:rana19} - The Rana19 TMO is an early deep-learning TMO that utilizes a Generative Adversarial Network (GAN) to create a fully-convolutional, parameter-free TMO. During training, the GAN objective was supplemented with perceptual losses to improve local detail and contrast retention.

\item Yang21 \cite{ref:yang21} - Yang21 is a recent deep-learning TMO that uses a deep convolutional neural network (CNN) to transform a multi-scale Laplacian pyramid decomposition of each input HDR frame, followed by an inverse Laplacian pyramid transform to reconstruct the SDR frame.

\item ITU21 \cite{ref:itu21} - ITU21 is a parameter-free TMO proposed by the ITU in Recommendation BT.2446 (``Approach A''). It uses a color-opponent representation of input HDR pixel values, followed by a non-linear transformation of the luma signal. The chroma channels are scaled proportionally with luminance.
\end{enumerate}

In addition to these methods, we also studied two other proprietary TMOs. The first is the DolbyVision TMO (DV), created by Dolby as part of the DolbyVision HDR standard, while the second is the Color Space Transform (CST), which is a popular gamut/tone-mapping tool used by colorists as part of the DaVinci Resolve video editing software.
\subsection{Using TMO Parameters to Vary Spatial Distortions}
\label{sec:database_spat_params}
When conducting TMO comparison/evaluation studies, the parameters of each TMO are usually optimized for each source content to yield the best possible appearance. As a result, the ``best-case'' performance of each TMO is compared. However, since the goal of our study is to understand tone-mapping distortions, we instead vary the spatial parameters of each TMO to generate videos that noticeably vary in spatial quality. In particular, we identify/introduce one spatial parameter of each TMO that we use to vary specific spatial properties of the tone-mapped videos. The list of spatial parameters identified for each TMO is presented below and summarized in Table \ref{tab:spat_params}

\begin{itemize}
    \item Hable, Reinhard02, Rana19, and Yang21 - We vary a parameter controlling \textbf{``desaturation''} when applying the TMO. The role of the desaturation parameter is to reduce color saturation, i.e., make colors appear ``more grey.'' Small amounts of desaturation may help correct oversaturated colors, but using large values of desaturation leads to almost colorless outputs. The values that we chose were \(\{0.0, 0.25, 0.5, 0.75\}\).
    \item Durand02 - We modify a \textbf{``base contrast''} parameter of Durand02, which controls the contrast of the base layer. Increasing the base contrast leads to higher global contrast. Care must be taken, since excessively large values of base contrast lead to losses of local contrast. The values we chose were \(\{10, 10^2, 10^3, 10^4\}\).
    \item Shan12 - The Haar wavelet basis is used to construct the multi-scale wavelet decomposition, and we varied the number of \textbf{wavelet levels} as the spatial parameter. Increasing the number of wavelet levels can improve local contrast enhancement, at the cost of a higher computational load. The values we selected were \(\{1, 2, 3, 4\}\).
    \item Reinhard12 - Since the Reinhard12 TMO is based on color appearance models, we varied the scene white point to simulate different \textbf{``viewing conditions''} and to introduce color distortions. The four viewing conditions used in this study were \{neutral, red, blue, green\}, where ``neutral'' refers to daylight illumination, and the other white points were designed to introduce red, blue, or green hues.
    \item Eilertsen15 - One of the key contributions of the Eilertsen15 TMO is the dynamic piecewise-linear tone curve designed to minimize contrast distortion. We varied the width of the piecewise-linear segments, called the \textbf{``segment width,''} to vary the coarseness of the tone curve. Using small segments reduces contrast distortion at the cost of a higher computational load. The values we chose were \(\{0.01, 0.03, 0.1, 0.3\}\).
    \item Oskarsson17 - Oskarsson17 is unique among the TMOs in this list due to its use of clustering to define a mapping between HDR and SDR values. Oskarsson17 clusters using implicit quantization of luminance and color. By varying the \textbf{number of clusters}, we trade computational complexity for finer quantization. Using too few clusters leads to visible banding and color quantization artifacts. The values we selected were \(\{8, 16, 32, 64\}\).
    \item ITU21 - ITU21 uses a \textbf{``peak luminance''} parameter that defines the maximum luminance of the grading HDR display, which is provided as metadata for PGC videos. By varying the peak HDR luminance, the global contrasts of the output videos are modified. The four values we selected were \(\{10^3, 10^4, 10^5, 10^6\}\) (nits).
\end{itemize}

It is important to re-emphasize that we are not comparing the capabilities or performances of the ten TMOs. Instead, we are empirically modeling the kinds of distortions that may occur when applying TMOs to video data, followed by standardized video compression. When creating these models, we follow our usual practice utilized in dozens of prior human studies, of creating a broad range of distortions (for each TMO) than might occur in practice, with the distorted outcomes of each TMO (and for each video content) being perceptually separable from each other. The goal being to model and capture the perceptual principles underlying human behavior with respect to the aggregate of distortions, rather than implementing any specific application-driven distortion parameters (in this case of commingled TMO + compression distortions). We have found that this approach yields ``better curve fitting,'' regardless of learning models' complexity, both over the wide range of training and testing distortions, as well as on any selected ``pragmatic'' range of distortions. In other words, learning the perceptual principles of distortions leads to better models that predict distortion.

\begin{table}[t]
    \centering
    \caption{TMO Parameters Used to Vary Spatial Distortions}
    \begin{tabular}{|c|c|c|}
        \hline
        TMO & Parameter & Values \\
        \hline
        Hable & Desaturation & \(\{0.0, 0.25, 0.5, 0.75\}\) \\
        Reinhard02 & Desaturation & \(\{0.0, 0.25, 0.5, 0.75\}\) \\
        Durand02 & Base Contrast & \(\{10, 10^2, 10^3, 10^4\}\) \\
        Shan12 & Wavelet Levels & \(\{1, 2, 3, 4\}\) \\
        Reinhard12 & Viewing Condition & \{Neutral, Red, Blue, Green\} \\
        Eilertsen15 & Segment Size & \(\{0.01, 0.03, 0.1, 0.3\}\) \\
        Oskarsson17 & Num. of Clusters & \(\{8, 16, 32, 64\}\) \\
        Rana19 & Desaturation & \(\{0.0, 0.25, 0.5, 0.75\}\) \\
        Yang21 & Desaturation & \(\{0.0, 0.25, 0.5, 0.75\}\) \\
        ITU21 & Peak luminance & \(\{10^3, 10^4, 10^5, 10^6\}\) \\
        \hline
    \end{tabular}
    \label{tab:spat_params}
\end{table}
\subsection{Using Temporal Modes to Vary Temporal Coherency}
\label{sec:database_temp_modes}
A key factor that successful video TMOs account for is the preservation of temporal coherency after tone mapping. Generally, TMOs may be modeled as non-linear mappings from HDR to SDR ranges that depends on aggregate statistics of input HDR frames, such as minimum, maximum, and mean luminance. Small temporal variations of these statistics, which may not affect the appearances of individual HDR frames, may lead to very noticeable temporal aberrations of the resulting sequences of tone-mapped SDR frames. These may manifest as flicker, blur, temporal shifts, and so on, i.e., temporal incoherencies of the tone-mapped SDR videos.

To study these distortions, we applied each of the TMOs to the source contents using three ``temporal modes'' - ``framewise,'' ``smoothed,'' and ``scene-level.'' As the name suggests, the \textbf{``framewise''} mode involves applying a TMO on each frame independently. This mode generally introduces the greatest degrees of temporal incoherency, which the other two modes aim to mitigate.

The \textbf{``smoothed''} temporal mode involves two passes of post-processing \cite{ref:boitard12} applied on a framewise tone-mapped SDR video. We use the ``relative brightness preservation'' (RBP) method, which uses a computed ``key value'' to quantify the brightness of each frame. RBP then seeks to maintain the key values of the tone-mapped SDR frames relative to the corresponding original HDR frames.

The key value of a frame \(I(i, j)\) of size \(M \times N\) is defined as
\begin{equation}
    \kappa = \exp\left(\frac{1}{MN} \sum_{i,j} \log\left(I(i,j) + \epsilon \right)\right).
\end{equation}
To apply RBP, first compute the key values of all HDR and framewise-tone-mapped SDR frames \(\kappa^{HDR}_f\) and \(\kappa^{SDR}_f\). Then, compute a scaling factor for each SDR frame:
\begin{equation}
    \alpha^{RBP}_f = \frac{\kappa^{HDR}_f}{\kappa^{HDR}_{max}} \frac{\kappa^{SDR}_{max}}{\kappa^{SDR}_f},
\end{equation}
using which the tone-mapped luminance is rescaled (multiplied). In this manner, the key values of the SDR frames, relative to the peak HDR key value, are made equal to the corresponding key values of HDR frames. The asuumption being, that preserving the smoothly varying key values of an HDR will promote temporal coherency of the tone-mapped SDR version of it. A key advantage of this method, along with its simplicity, is that it can be applied to any TMO applied in a first pass, including the proprietary DV and CST TMOs.

The primary reason for seeking temporal coherence is that by suitably estimating critical parameters at the frame level, which vary across frames, the outcomes of TMOs can be made more temporally satisfying. To further address this, we developed a two-pass \textbf{``scene-level''} temporal mode. To apply a TMO in the scene-level mode, we first identify salient parameters that are estimated on each frame. Then, as described in the following, we reconfigure the TMO to estimate the same parameters at the scene (rather than frame) level in the first pass. During the second pass, these parameters, which are now constant across frames, are used to tone-map each frame.

When applying this method to a video, it is important to first detect scene changes, then re-estimate the parameters on each scene. However, since our database contains short clips that only consist of one scene, we estimate the scene-level parameters once per clip.

To illustrate scene-level tone-mapping, we describe an application example using the Hable TMO. The non-linear tone curve used by the Hable TMO is given by
\begin{equation}
    I_{SDR}(i,j) = \frac{g_{Hable}\left(I\left(i, j\right)\right)}{g_{Hable}\left(I_{max}\right)},
\end{equation}
where \(g_\textit{Hable}\) is of the form
\begin{equation}
    g_\textit{Hable}(x) = \frac{Ax^2 + Bx + C}{Ax^2 + Dx + E} - F
\end{equation}
and \(I_{max}\) is estimated as the maximum frame luminance by default. When adapting Hable to the scene-level mode, we estimate \(I_{max}\) as the maximum luminance over all frames, then use the same value when tone-mapping all frames. A similar approach is used to adapt all of the ten open-source TMOs, with details regarding the parameters adapted in each case being provided in Table \ref{tab:temp_params}.

\begin{table}[t]
    \centering
    \caption{TMO Parameters Adapted to Scene-Level Tone-Mapping}
    \begin{tabular}{|c|c|}
        \hline
        TMO & Parameter(s) Adapted to Scene-Level \\
        \hline
        Hable & Scene max. luminance\\
        Reinhard02 & Scene key value \\
        Durand02 & Scene min. \& max. luminance\\
        Shan12 & Scene min. \& max. luminance\\
        Reinhard12 & Scene \& view max. neural response, view semi-saturation\\
        Eilertsen15 & Piecewise-linear local tone curves\\
        Oskarsson17 & Log-luminance histogram used to create clusters. \\
        Rana19 & Scene min. \& max. luminance\\
        Yang21 & Scene min. \& max. log-luminance\\
        ITU21 & Scene mean log-luminance\\
        \hline
    \end{tabular}
    \label{tab:temp_params}
\end{table}
\subsection{Lossy Compression}
\label{sec:database_compression}
The bandwidth requirements of HD video streaming over the Internet necessitate lossy compression. Quite a few publicly available subjective databases have been created to study the effects of compression on visual quality, both for SDR \cite{ref:vqeghd3} \cite{ref:livestream} and HDR \cite{ref:live_hdr} videos.

Due to the ubiquity of video compression in streaming, we aimed to study the combined and interacting effects of tone-mapping and compression. For example, consider a TMO that boosts local contrast. Such a TMO may yield videos that are less compressible, causing steeper losses of visual quality from compression. Conversely, a TMO that does not boost local contrast may achieve a lower quality at high bitrates, but may also suffer less from compression.

\begin{figure}[t]
    \centering
    \includegraphics[width=0.8\linewidth]{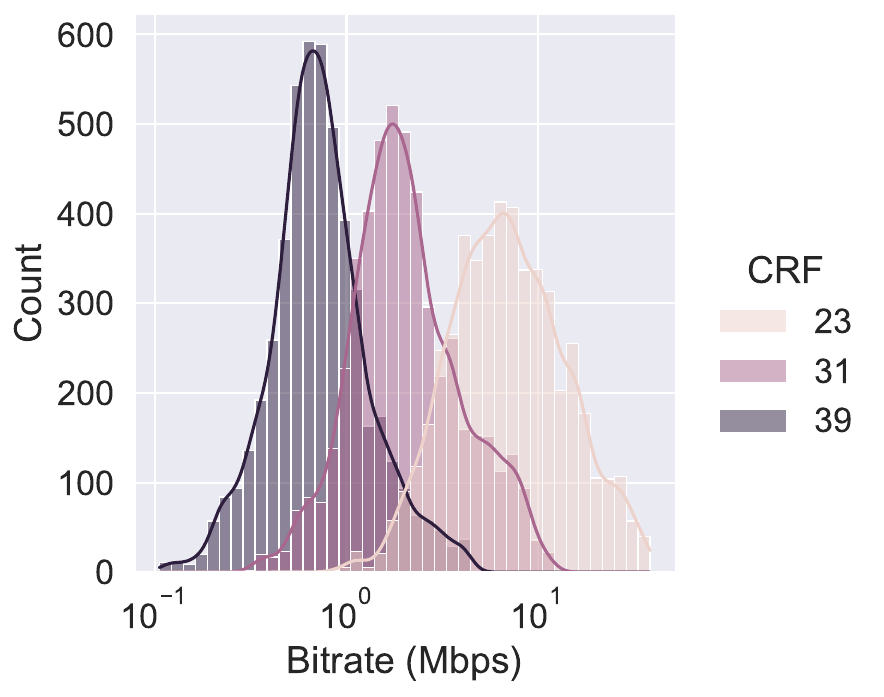}
    \caption{Distributions of bitrates of compressed tone-mapped videos.}
    \label{fig:bitrate_distributions}
\end{figure}

To study these issues, we compressed all of the experimental tone-mapped videos at three compression levels using the libx264 encoder \cite{ref:x264} in the Constant Rate Factor (CRF) encoding mode. We selected H.264 instead of HEVC because of its much wider device support and current user base. For example, according to data from ``Can I Use,'' only 20.57\% of users experience full HEVC support, with another 70.29\% experiencing partial support based on hardware requirements \cite{ref:can_i_use_hevc}. By contrast, 98.1\% of all users experience full H.264 support \cite{ref:can_i_use_avc}. This gap is partly because of the greater computational demands of HEVC and licensing difficulties. The three compression levels selected were CRF 23, 31, and 39, again with the aim of creating a wide range of perceptually separable compression distortions. The distribution of bitrates corresponding to the three compression levels is shown in Fig. \ref{fig:bitrate_distributions}. The significant overlap of the bitrate distributions between the CRFs demonstrates the wide range of content complexities, and the non-monotonic relations of distortion with content (because of masking \cite{ref:masking1} \cite{ref:masking2}) among the videos in the LIVE-TM-HDR database.

\section{Crowdsourced Subjective Quality Assessment}
\label{sec:study}
To obtain a large number of ground-truth visual quality scores on the tone-mapped videos in the LIVE-TM-HDR database, we conducted an online subjective study whereby we recruited naive human subjects to view and rate all of the tone-mapped and compressed SDR videos. Since the appearances of HDR and SDR videos are quite sensitive to display settings and ambient viewing conditions, a typical approach to conducting HDR subjective studies usually involves setting up a controlled environment and inviting subjects to undertake the study in person. 

However, because of the (intentionally) large size of the LIVE-TMHDR database, and our desire to collect many human annotations of each video, an in person study was infeasible. Moreover, we also reasoned that tone-mapped videos streamed over the internet will be viewed by consumers in home conditions that may vary significantly between users. In these scenarios, much of the viewing and display information may be hidden from the server, or may not be easily incorporated into a tone-mapping process. So, while ``uncontrolled' human viewing is a more challenging test setup, it is also a more realistic model of actual viewing.

Therefore, we obtained ground-truth subjective quality scores by conducting an online large-scale crowd-sourced subjective study on the Amazon Mechanical Turk (AMT) \cite{ref:amt} platform, which has been successfully used by ourselves and others in prior studies of crowdsourced subjective quality assessment \cite{ref:live_wild} \cite{ref:kadid_10k} \cite{ref:live_vqc} \cite{ref:patch_vq}. Before describing our execution of this large project, we describe the design and outcome of a small-scale pilot study that we conducted prior to the crowdsourced study to validate our overall protocol. We also lay out the organization of the overall database into rating sessions, and the design of the crowdsourced subjective rating protocol.

\subsection{The Pilot Study}
\label{sec:study_pilot}
Due to the relatively novel nature of video tone-mapping distortions on challenging HDR contents, we thought it possible that subjects might not agree on the ``true quality'' of tone-mapped videos. In other words, that the quality labeling task could prove to be difficult and noisy. Tone-mapping involves making subjective decisions regarding more complex and subtle aspects of distorted videos, such as balances between local and global contrast, crushing or losses of extreme bright/dark regions, and editing of color to appear natural in a reduced brightness range. Indeed, the design of tone mapping operators involves artistic and aesthetic elements, as well as preserving technical quality. Therefore, despite carefully applying varying degrees of distortion, a study may prompt subjects to perceive all of the displayed videos to be of low quality (``everything is distorted, and without aesthetics'') or high quality (``everything is aesthetically pleasing, and nothing is distorted''). These would be contrary to the goals of our study, which is to model the perception of a wide range of visual qualities. 

Therefore, we decided to validate the study protocol locally before proceeding to crowdsource it at large scales. We sought to examine two critical aspects of the likely outcome of our experiment: inter-subject correlation, i.e., the degree to which the human subjects agree with one another regarding the quality of each video; and the quality diversity of the displayed videos as reported by human subjects. Thus, we conducted a small-scale pilot study by recruiting fifteen volunteer college students.

To set up the pilot study, we extracted a representative set of 320 videos, from the dataset of 15,000, by randomly sampling eight tone-mapped and compressed videos for each of the 40 source contents. Since the expert tone-mapped videos are significantly underrepresented, yet critical to the database, care was taken to ensure that all of them were included in the pilot subset. The 320 videos were then split into two ``batches'' of 160 videos each, with each batch containing 10 PGC and 10 UGC videos.

The batches were presented to each of the subjects in two sessions of roughly 45 minutes each. To simulate the large-scale AMT study, the same interface and instructions were used in the pilot study. Moreover, the subjects participated in the study from their own uncontrolled home settings (in Austin) using their own display devices. The results of the pilot study are presented in Section \ref{sec:data_analysis}. Since we recruited more reliable participants, the ratings obtained from the pilot study were also used to provide ``gold scores'' used for subject rejection in the large-scale online study. The use of gold scores to verify crowdsourced subjective ratings is explained further in Section \ref{sec:study_amt}.
\subsection{Partitioning Database into Representative Batches}
Due to the large size of the video database, we partitioned it into small batches that could viewed and rated by subjects in sessions of reasonable durations. This is a feature of all crowdsourced subjective rating experiments, and the typical strategy is to randomly shuffle the set of all videos, then partition them into batches of appropriate sizes. 

 For unstructured databases such as LSVQ \cite{ref:patch_vq} and KonViD-150k \cite{ref:konvid_150k}, each test video presented to a given subject contains unique content. Hence, random sampling into batches automatically ensures that they are representative. However, because of the structured nature of the LIVE-TMHDR database, random sampling could lead to some source contents being over/under-represented in any particular batch.
 
Therefore, we would like to devise a structured sampling scheme that ensures that all source contents present in a batch were represented equally. Due to the limited size of each batch, there is a tradeoff between the number of unique contents and the number of test videos per content. To balance this tradeoff, we include ten unique source contents in each batch.

\begin{figure}[b]
    \centering
    \includegraphics[width=\linewidth]{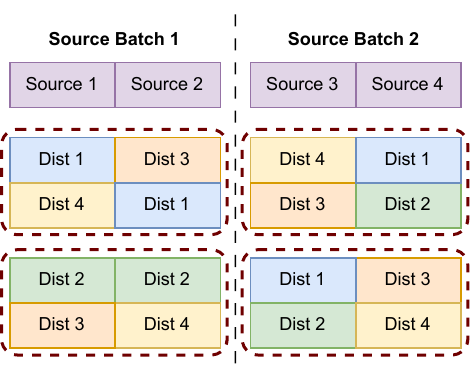}
    \caption{Example of a structured sampling scheme yielding four representative batches of four videos a piece.}
    \label{fig:structured_sampling}
\end{figure}

To achieve this, we first shuffled the set of tone-mapped videos corresponding to each content separately, yielding 40 sets of 375 videos each. We then created four ``source batches'' by randomly partitioning the 40 source contents into four batches of ten each. Care was taken during this process to place five unique UGC and PGC source contents in each source batch, ensuring that each subject would see an equal number of UGC and PGC contents. Each batch was finally created by drawing an average of 7.5 videos from each of the ten source contents per source batch, yielding 75 test videos per batch. This procedure was used to create 200 batches that partitioned the overall database of 15,000 videos. An example of a structured sampling scheme to partition a database containing four source contents and four distortions per source into four batches is illustrated in Fig. \ref{fig:structured_sampling}.

Since most of the subjects participated in a small number of sessions relative to the size of the full database, we sought to eliminate any biases by ensuring that each batch was ``representative'' of the LIVE-TMHDR database. We defined a batch to be representative if the distribution of distortions in the batch was similar to those in the entire database.

To analyze this, we measured how uniformly each TMO, spatial parameter, temporal mode, and compression level was distributed among the test videos in each batch. Since the LIVE-TM-HDR database has an equal number of each of these categories, the uniformity of the distribution of distortion parameters is analogous to how ``representative'' each batch is. For example, a batch that contains 75 videos all tone-mapped using the ``framewise'' temporal mode is a worst-case example of highly non-uniform, and hence is not representative of the dataset, in this case relative to the distributions of temporal modes. On the other hand, a batch containing an equal number of videos tone-mapped using each of the three temporal modes would be representative of the database.

We subsequently quantified the representativeness of a batch, i.e., the uniformity of the distribution of the distortion parameters, using normalized Shannon entropy. For example, to measure how representative a batch is with respect to the (\(N=3\)) temporal modes, we obtained the relative frequency of each mode (say, \(f_i\)) among the test videos in that batch. We then computed its representativeness as

\begin{equation}
    R = \frac{1}{\log(N)}\sum\limits_{i=1}^{N}{f_i\log(1/f_i)}.
\end{equation}

\begin{figure}
    \centering
    \includegraphics[width=\linewidth]{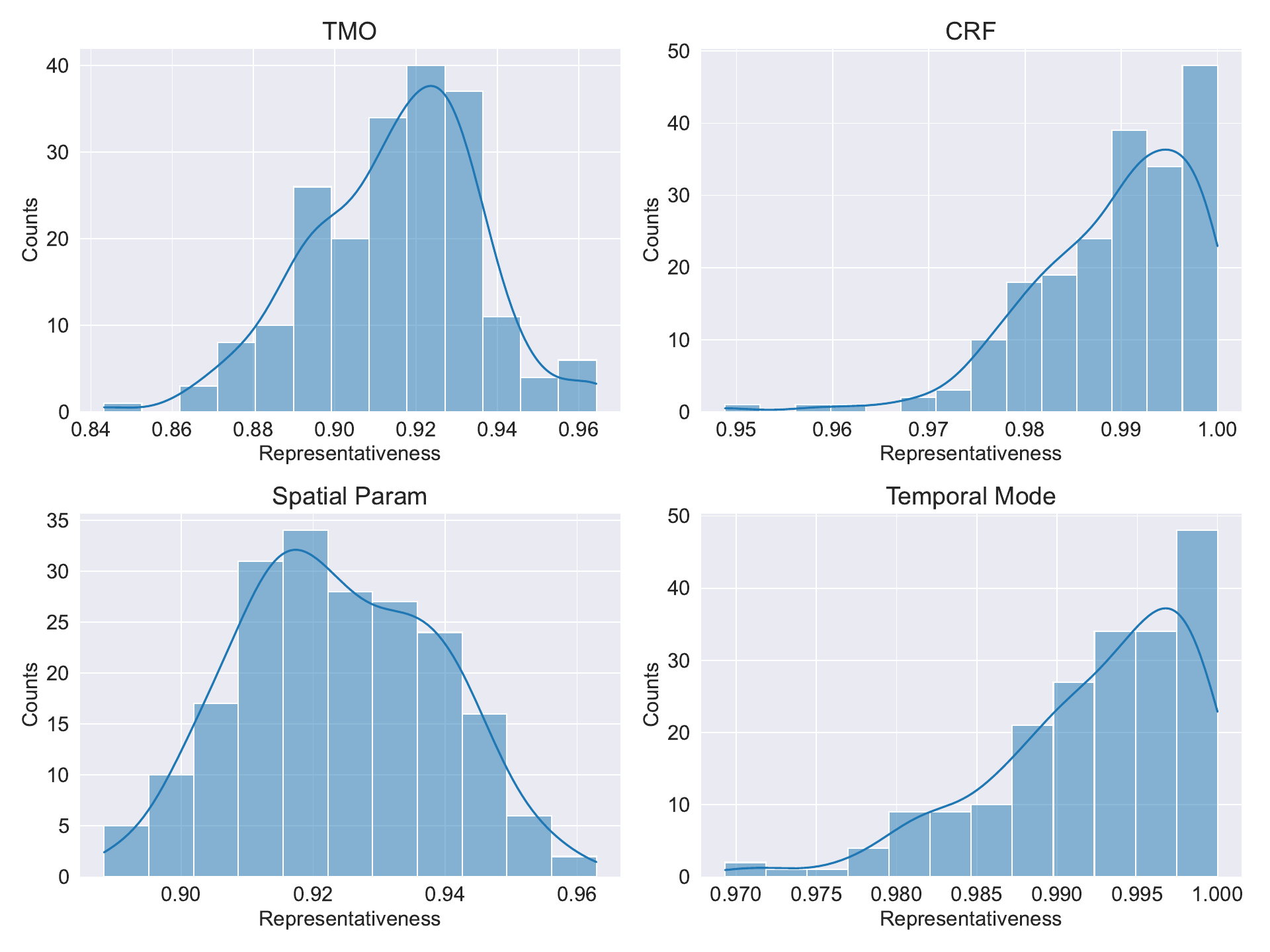}
    \caption{Distributions of representativeness of 200 batches.}
    \label{fig:representativeness}
\end{figure}

The distribution of the batch representativeness values, for each of the four distortion parameters, across the 200 test batches, is shown in Fig. \ref{fig:representativeness}. From the figure, it may be seen that all of the batches proved to be highly representative, in terms of TMOs, and their spatial parameters, temporal modes, and compression levels.

\subsection{The Large-Scale Crowdsourced Subjective Study}
\label{sec:study_amt}
After conducting the pilot study and partitioning the database into 200 batches, we proceeded to design and execution of the the main large-scale subjective study on the AMT platform. We followed a similar methodology as prior work \cite{ref:live_vqc} \cite{ref:patch_vq} and presented each batch as a Human Intelligence Task (HIT) to workers. 

The study was conducted in three phases A to C. In phase A, the study was restricted to ``highly-qualified'' workers until eighteen ratings were obtained per HIT. Similar to the criteria used in \cite{ref:patch_vq}, we selected ``highly-qualified'' workers as those having \(>\)90\% lifetime HIT acceptance rate and \(>\)10,000 lifetime HITs accepted. Phase B was opened to a wider pool of ``regular'' workers until 27 more ratings were obtained for each HIT. ``Regular'' workers were selected as those having \(>\)75\% lifetime HIT acceptance rate and \(>\)1,000 lifetime HITs. Moreover, all workers were limited to using laptops, desktops, or TV displays by detecting and rejecting workers who were using mobile phones or tablet devices.

Finally, the number of ratings for each HIT in phase C was decided after performing subject rejection for phases A and B, to obtain an average of 50 ratings per HIT over all HITs and phases. Partitioning the study into phases in this manner enables fault tolerance by isolating potential errors to small subsets of the data. However, no such errors were encountered in this study.

\begin{figure*}[t]
    \centering
    \includegraphics[width=0.8\linewidth]{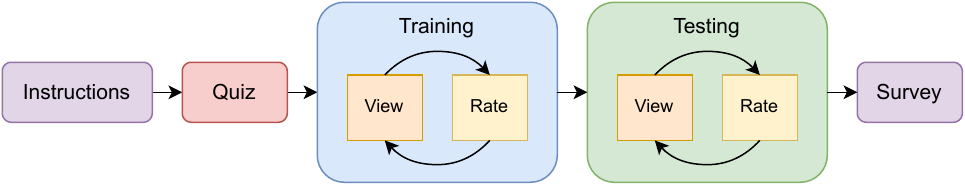}
    \caption{Flowchart of a typical HIT on the AMT platform.}
    \label{fig:amt_flowchart}
\end{figure*}

\begin{figure*}[t]
    \centering
    \includegraphics[width=0.9\linewidth]{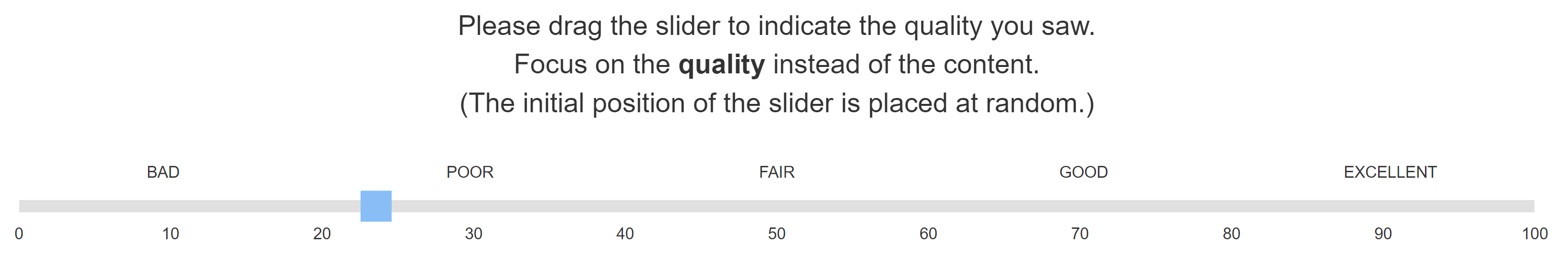}
    \caption{The rating slider used to provide quality ratings.}
    \label{fig:slider}
\end{figure*}

A flowchart illustrating the workflow of a typical rating session is presented in Fig. \ref{fig:amt_flowchart}. The stages are summarized:
\begin{itemize}
    \item \textbf{Instructions}: The subject was presented with general instructions regarding the goal of the study, and specifically instructed to rate quality, rather than aesthetics (such as framing or composition) or content. Sample videos of broad quality categories (Bad, Poor, Fair, Good, Excellent) were provided without explanation, to avoid training subjects to search for specific distortions. These sample videos were not included in the LIVE-TMHDR database to avoid bias. The ethics policy was shared, establishing the expectation of rating videos earnestly and with proper attention.
    \item \textbf{Quiz}: A short quiz was administered, with the goal of ensuring that subjects had read and understood the instructions. Subjects had to answer at least five out of six questions correctly to proceed. Subjects who failed the quiz were redirected to the instructions. 
    \item \textbf{Training}: Subjects were trained to use the rating interface by rating four videos. These videos were generated using source contents that were not included in the database, to avoid any bias. After each video was played, the rating slider shown in Fig. \ref{fig:slider} was presented. Subjects could choose any point on the continuous scale, and the recorded score was scaled to the range 0-100 and rounded to the nearest integer.
    \item \textbf{Testing}: The testing phase was the main phase of the session, whereby videos were presented in a randomized order to the subject for rating. Each video could be viewed only once, and a rating had to be provided to proceed to the next video. The rating procedure was identical to that in the training session, and subjects' progress was periodically reported on the screen as a percentage of the total videos rated.
    \item \textbf{Survey}: Subjects were requested to undertake an exit survey to collect study and demographic information such as the display device (restricted to TVs, desktops, and laptops), age, familiarity with HDR videos, etc. 
\end{itemize}

\subsection{Rejecting Unreliable Subjects}
While a vast majority of subjects on the AMT platform provide genuine quality ratings to test videos, some subjects may provide unreliable ratings either consciously or due to extraneous circumstances. Due to the monetary compensation provided upon completion, some workers may feel incentivized to provide random or meaningless ratings to videos. Even workers who intend to participate earnestly may experience disruptions during the study which unduly affect the ratings they provide. Therefore, to preserve data quality, we utilize various criteria that are applied during and after the study session to identify and exclude unreliable subjects. These criteria and techniques are summarized:

\begin{itemize}
    \item \textbf{Device}: To maintain a degree of device uniformity across viewing conditions, we detected whether mobile devices were being used by the subjects, by examining browser metadata. Such subjects detected to be using mobile devices were disallowed from participating in the study.
    \item \textbf{Browser Resolution}: Workers with displays having a resolution lower than 1280\(\times\)720 were not allowed to participate to ensure that low browser resolution did not introduce aliasing artifacts that could significantly affect the quality of videos. In addition, the browser zoom was set to 100\% to maintain the desired viewing condition.
    \item \textbf{Playback Time}: Though every video was loaded fully before playback, hardware resource constraints such as memory and processing power at the worker's end could lead to stalls, which affect quality. If over 50\% of all videos suffered from significant stalls, the subject was rejected. Moreover, any attempts to speed up the video during the study led to immediate termination of the session and rejection.
    \item \textbf{Repeated Videos}: To measure subject reliability, six randomly selected videos were repeated twice during the study. If the difference between the two scores of repeated videos was over 20 for more than 3/6 videos, the HIT was deemed unreliable and rejected.
    \item \textbf{Golden Videos}: To measure subjects' understanding of the study, four ``golden videos'' were presented in each session, for which quality values are already known. Typically, such videos are selected from previous crowdsourced studies of the same kind. However, since LIVE-TMHDR is the first database of its kind, we instead selected four videos from the database of videos used in the pilot study. Care was taken to ensure that none of the gold videos were included among the 75 videos to be rated. The mean opinion score obtained from the pilot study was used as a reference, and subjects whose ratings deviated by more than 30 from the reference value for at least three out of four golden videos were rejected.
    \item \textbf{Random/Meaningless Scores}: Two common strategies used by insincere participants are to provide the same score to every video or to nudge the rating slider before submissions. We detected  these cases by analyzing the variation in scores and the deviation between the initial and final positions of the rating slider. Specifically, if the standard deviation of ratings or the average difference between the initial and final slider positions was less than five, the subject was deemed to have input meaningless scores. Subjects found to engage in either of these strategies were rejected and blocked.
\end{itemize}

In addition to using the above conditions to flag submissions and subjects for rejection, all flagged submissions were reviewed manually prior to rejection. In total, using the aforementioned criteria, 1287 submissions were rejected, which accounted for almost 11.3\% of all submissions.

\section{Processing and Analyzing Subjective Data}
\label{sec:data_analysis}

\subsection{Obtaining Quality Labels from Subject Ratings}
\label{sec:labeling}
Using the subjective methodology described above, we obtained a total of over 750,000 subjective opinions from over 1,600 unique subjects, with an average of 50.43 ratings per test video. The ratings obtained in this manner were then processed to obtain a single quality label per video. The simple method that is most widely used is to compute the mean opinion score (MOS) as the average of the subjective ratings obtained on the video:
\begin{equation}
    MOS_v = \frac{1}{N_v}\sum\limits_s r_{sv},
\end{equation}
where \(r_{sv}\) is the rating provided by subject \(s\) to video \(v\) and \(N_v\) is the total number of ratings obtained by video \(v\).

\begin{figure}[t]
    \centering
    \includegraphics[width=\linewidth]{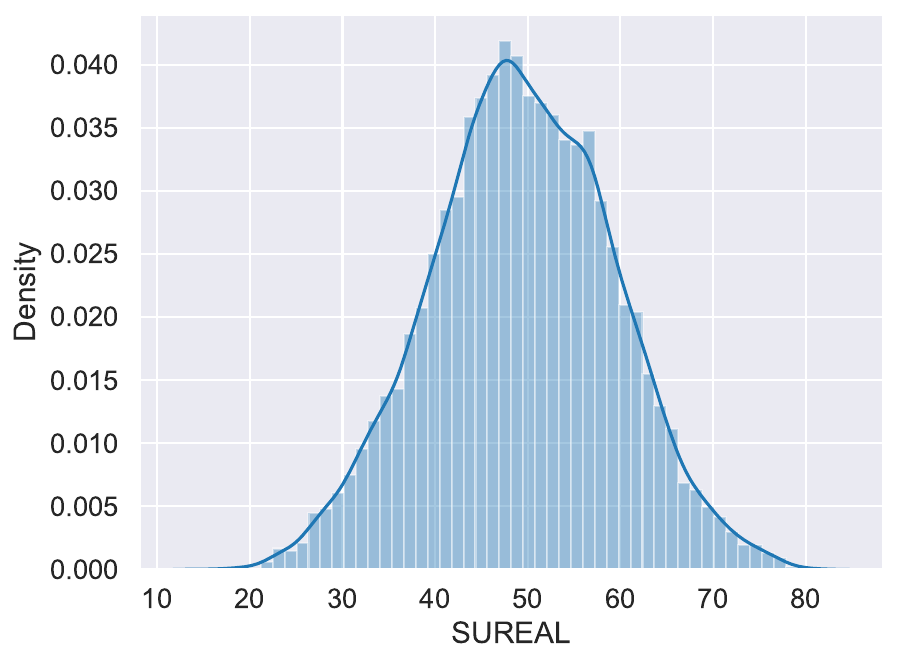}
    \caption{Histogram of quality scores obtained using SUREAL.}
    \label{fig:sureal_hist}
\end{figure}

However, SUREAL \cite{ref:sureal} is a recent and more sophisticated approach to obtaining quality labels, which involves computing Maximum Likelihood Estimates of the ``true quality,'' assuming the following subject rating model
\begin{equation}
    R_{sv} = Q_{v} + b_s + \sigma_s \mathcal{N}(0, 1)
\end{equation}
where \(Q_v\) is the ``true quality'' of video \(v\), \(\mathcal{N}(0, 1)\) is a standard Gaussian random variable, \(b_s\) denotes ``subject bias,'' and \(\sigma_s\), which denotes ``subject variability.'' The Alternating Projection (AP) solver described in \cite{ref:sureal} is used to estimate the model parameters. The distribution of estimated SUREAL scores for all videos from the LIVE-TM-HDR database is presented in Fig. \ref{fig:sureal_hist}. From the figure, it may be seen that the database spans a wide range of qualities, from a score of 20 at the lower end to 80 at the higher end.

\subsection{Inter-Subject Correlation Analysis}
\label{sec:data_analysis_inter_subj}
\begin{figure}[ht]
    \centering
    \includegraphics[width=0.9\linewidth]{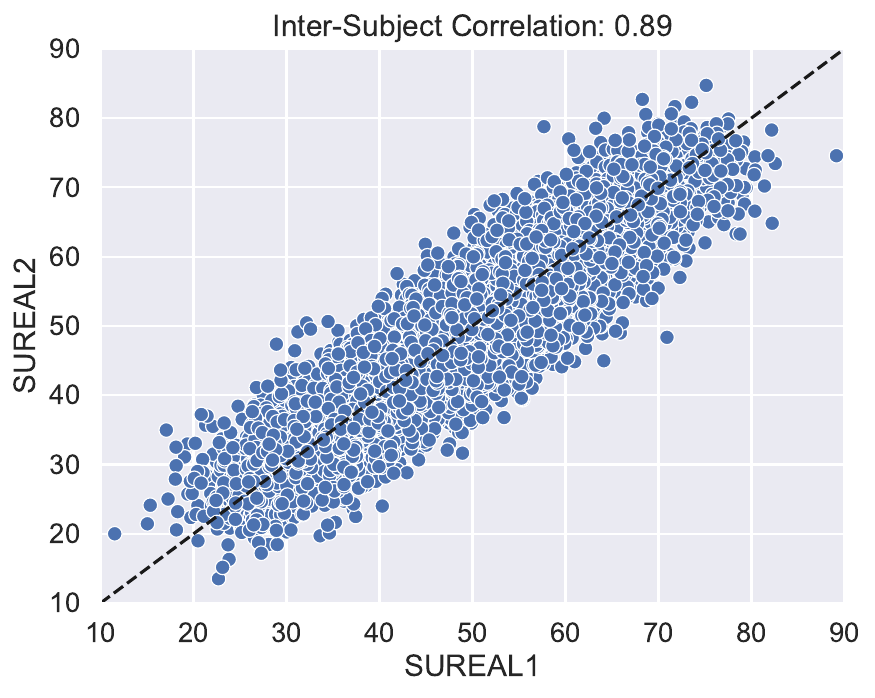}
    \caption{Scatter plots comparing quality estimates obtained from random partitions of subjective ratings.}
    \label{fig:inter_subj}
\end{figure}

To understand the reliability of subject ratings, we evaluated inter-subject correlation by randomly splitting the set of ratings for each video into two subsets of equal size. SUREAL scores were estimated independently for the two scores, yielding two quality labels per video. The Spearman's Rank Order Correlation Coefficient (SROCC) between these labels was measured, and the entire procedure was repeated 50 times. The average SROCC value obtained over 50 iterations is a measure of the inter-subject correlation, where a high value indicates a high degree of agreement between subject ratings. In our experiments, the measured inter-subject correlation was 0.89. A sample scatter plot corresponding to a random split of subjective ratings is shown in Fig. \ref{fig:inter_subj}. 

\subsection{Comparing the Pilot and Crowdsourced Studies}
\label{sec:data_analysis_main_v_pilot}
\begin{figure}[b]
    \centering
    \includegraphics[width=\linewidth]{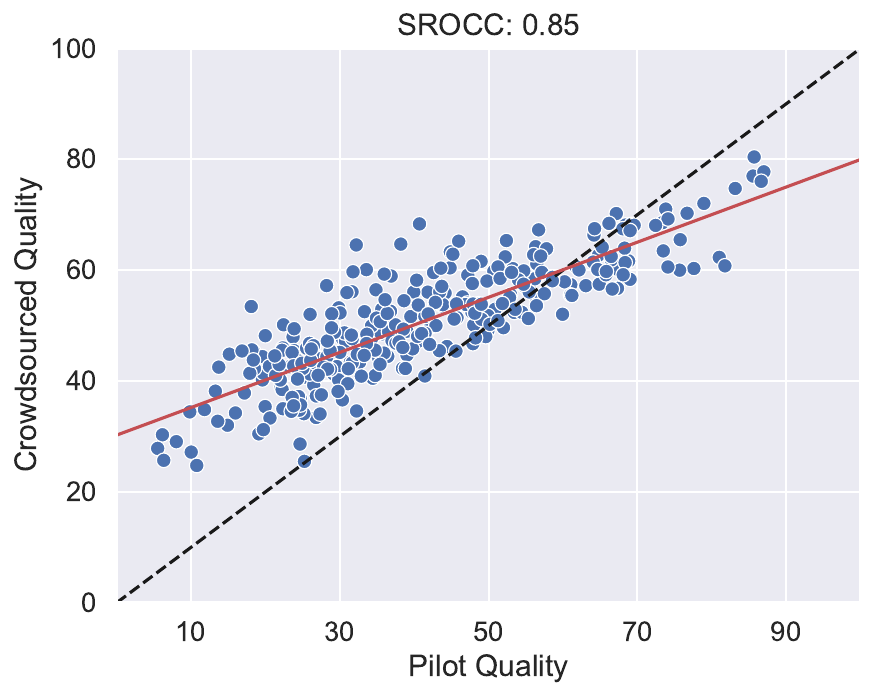}
    \caption{Comparing ratings from the pilot and crowdsourced studies.}
    \label{fig:main_v_pilot}
\end{figure}
The inter-subject analysis presented above demonstrates self-consistency among participants of the crowdsourced study. However, we are also interested in verifying the ``accuracy'' of these quality ratings. Small-scale studies using trusted participants typically provide high-quality ratings at the cost of scalability, while large-scale crowdsourcing is more scalable at the cost of relying on unknown subjects. Due to our reliance on crowdsourcing, we compared the two rating protocols to verify its veracity as a data-collection method. 

The data obtained from the pilot study is an ideal test-bed for such analysis. We compute independent SUREAL quality estimates for the 320 pilot videos from ratings obtained in both the pilot and the crowdsourced studies. A scatter plot showing pairs of quality estimates for each video is shown in Fig. \ref{fig:main_v_pilot}, and the correlation between the two was found to be 0.85.

An interesting phenomenon that may be observed from the figure is that participants in the pilot study generally rated videos as having lower overall quality as compared to the crowdsourced study. In particular, they rated low-quality videos lower and high-quality videos higher. One reason for this may be that, while most participants in the crowdsourced study are likely to be completely naive subjects, the university students recruited for the pilot study were familiar with general video distortions. This may have led them to be able to more closely discern distortions in the lower quality range, and identify improvements in quality in the higher range. This may be an important phenomenon to account for when translating the results of laboratory subjective studies to real-world applications.
\subsection{Analyzing the Impact of TMOs on Quality}
Due to the diversity in the set of TMOs included in the study, each TMO imparts a unique combination of distortions on the source HDR videos. Therefore, we would like to understand the impact of each TMO on video quality. This analysis may also be considered a direct comparison of various TMOs, though that is not the primary purpose of this study.

\begin{figure}[t]
    \centering
    \includegraphics[width=\linewidth]{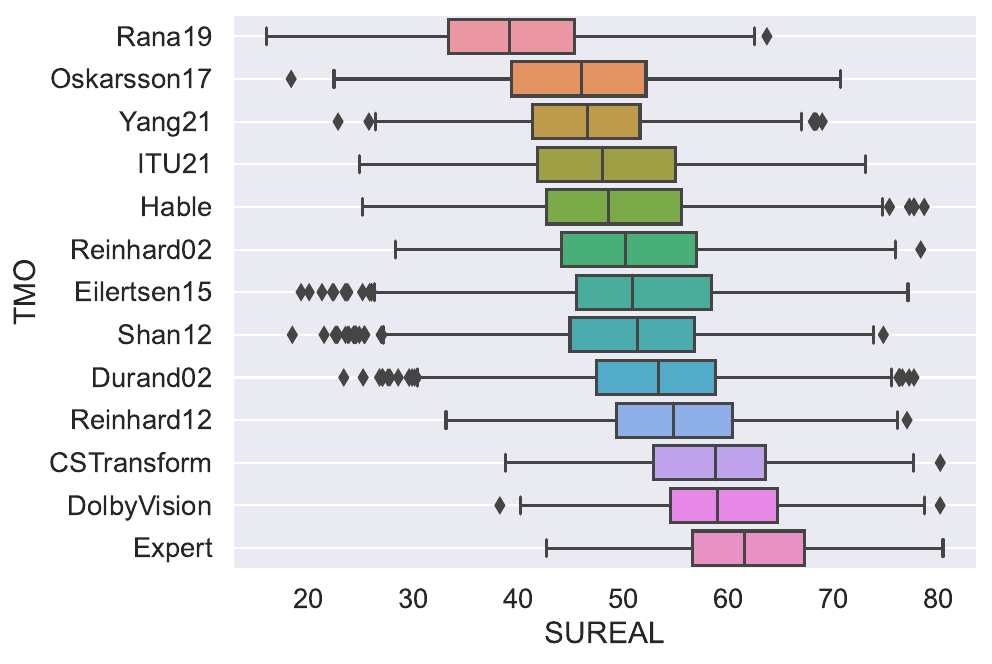}
    \caption{Distributions of SUREAL scores by TMO.}
    \label{fig:tmo_sureal}
\end{figure}

In Fig. \ref{fig:tmo_sureal}, we present box plots showing the distribution of quality scores for videos tone-mapped using each of the 13 TMOs. It may be observed that as expected, Expert tone-mapping achieved the highest average quality, followed by the proprietary DolbyVision and Color Space Transform algorithms. 

\begin{figure}[t]
    \centering
    \includegraphics[width=0.9\linewidth]{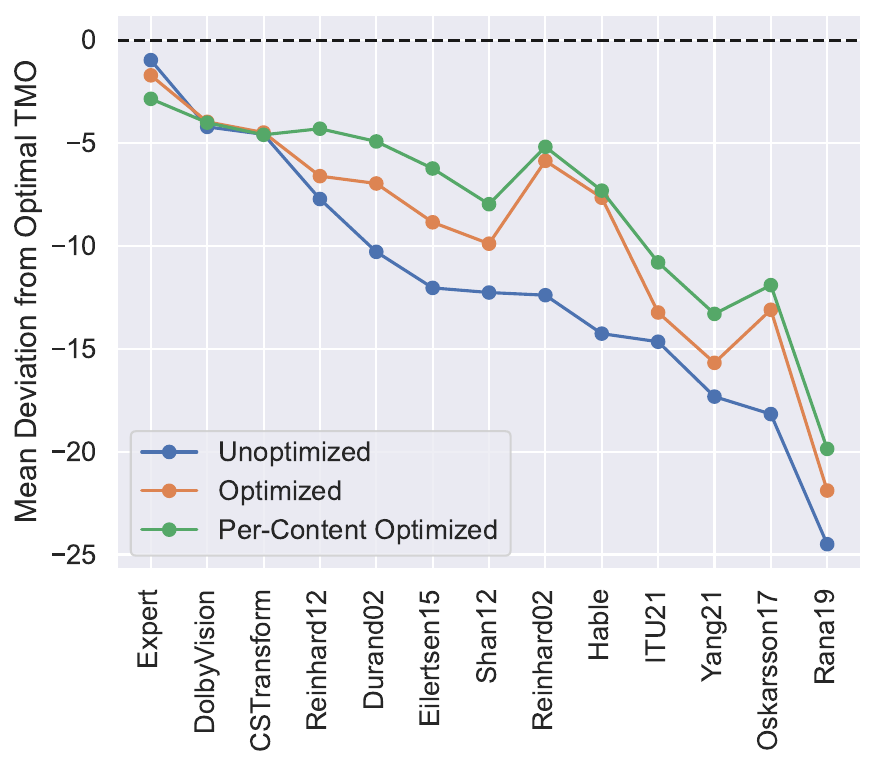}
    \caption{Mean deviation from optimality for each TMO.}
    \label{fig:opt_deviation}
\end{figure}

\begin{table}[b]
    \centering
    \caption{Best ``Average-Case'' Spatial and Temporal Parameters of Each TMO Among Those Considered in the Study}
    \label{tab:opt_params}
    \begin{tabular}{|c|c|c|}
        \hline
        TMO & Best Spatial Parameter & Best Temporal Mode \\
        \hline
        Hable & Desaturation = 0.0 & Smoothened \\
        Reinhard02 & Desaturation = 0.0 & Scene-wise \\
        Durand02 & Base Contrast = \(10^2\) & Smoothened \\
        Shan12 & Wavelet Levels = 1 & Smoothened \\
        Reinhard12 & Viewing Condition = Neutral & Smoothened \\
        Eilertsen15 & Segment Size = 0.3 & Scene-wise \\
        Oskarsson17 & Num. of Clusters = 64 & Smoothened \\
        Rana19 & Desaturation = 0.0 & Scene-wise \\
        Yang21 & Desaturation = 0.25 & Smoothened \\
        ITU21 & Peak luminance = \(10^4\) & Framewise \\
        \hline
    \end{tabular}
\end{table}
Since the source contents vary in terms of their spatial and temporal complexities and range of brightness and colorfulness values, one may expect that the ``best TMO'' for each content might be different. However, practical HDR processing pipelines would typically use one TMO for all contents. So, for each TMO investigated in the study, we quantified the ``deviation from optimality'' averaged over all contents. 

We conducted this analysis in three different ways. First, we considered ``unoptimized'' TMOs, where the quality delivered by a TMO on a content (\(c\)) was averaged over all choices of parameter values. Theoretically, this may be interpreted as the expected quality when choosing parameters randomly. Then, we defined the mean deviation from optimality (MDO) for a TMO \(T\) having parameters \(\theta_\textit{T}\) as
\begin{equation}
    \textit{MDO}_\textit{Avg}(T) = E_c\left[Q_\textit{Avg}(c, T) - \max_T Q_\textit{Avg}(c, T)\right],
\end{equation}
where \(E[\cdot]\) denotes averaging and \(Q(c, T)\) is the average quality when content \(c\) is tone-mapped using TMO \(T\):
\begin{equation}
    Q_\textit{Avg}(c, T) = E_{\theta_T} \left[ Q\left(c, T; \theta_T\right) \right].
\end{equation}

A better approach to tone-mapping involves ``optimizing'' TMOs by selecting a common set of ``best'' parameters to apply to all contents. We obtained the best average-case parameters by optimizing for mean quality over all contents in the database. The optimal parameters for the ten open-source TMOs are shown in Table \ref{tab:opt_params}. Then, the MDO for the optimized TMOs is defined as
\begin{equation}
    \textit{MDO}_\textit{Opt}(T) =  E_c\left[Q_\textit{Opt}\left(c, T\right) - \max_T Q_\textit{Opt}\left(c, T\right)\right],
\end{equation}
where the optimized quality is
\begin{equation}
   Q_{Opt}(c, T) = Q\left(c, T; \theta_{T, Opt}\right).
\end{equation}

Finally, the best possible approach to tone-mapping would be to separately identify the best parameter values for each content, leading to ``per-content-optimized'' TMOs. The MDO for per-content-optimized TMOs is then defined as
\begin{equation}
    \textit{MDO}_\textit{PCOpt}\left(T\right) =  E_c\left[Q_\textit{PCOpt}\left(c, T\right) - \max_T Q_\textit{PCOpt}\left(c, T\right)\right],
\end{equation}
where the optimized quality is
\begin{equation}
   Q_\textit{PCOpt}(c, T) = \max_{\theta_T}Q\left(c, T; \theta_T\right).
\end{equation}

Fig. \ref{fig:opt_deviation} plots the MDO of the thirteen TMOs, for all three levels of optimization. The first observation that may be made is that Expert tone-mapping nearly always achieved the best quality. The next best TMOs were the proprietary TMOs. However, interestingly, when parameters were optimized, even simple TMOs such as Reinhard02 and Durand02 could achieve similar quality. Finally, optimizing parameters on a per-content basis led to further improvements in quality.

Note that in this analysis, the best parameters were identified using subjective quality scores obtained from the crowd-sourced study. However, to achieve per-content optimization in practice, a predictive model of subjective quality would be required. The output of the model could then be used in place of subjective quality scores to automate tone-mapping decisions in production.

\begin{figure*}
    \centering
    \subfloat[Effect of desaturation.\label{fig:desat_mdo}]{%
      \includegraphics[width=0.45\linewidth]{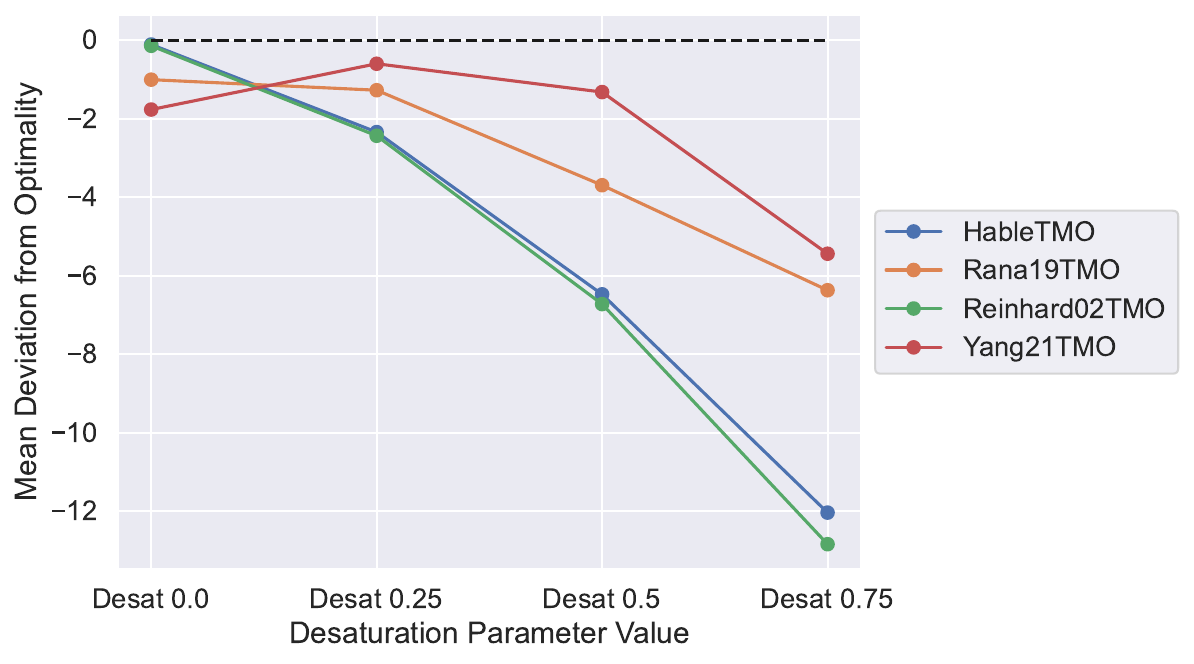}
    }
    \hfill
    \subfloat[Effect of temporal modes.\label{fig:temp_mdo}]{%
      \includegraphics[width=0.48\linewidth]{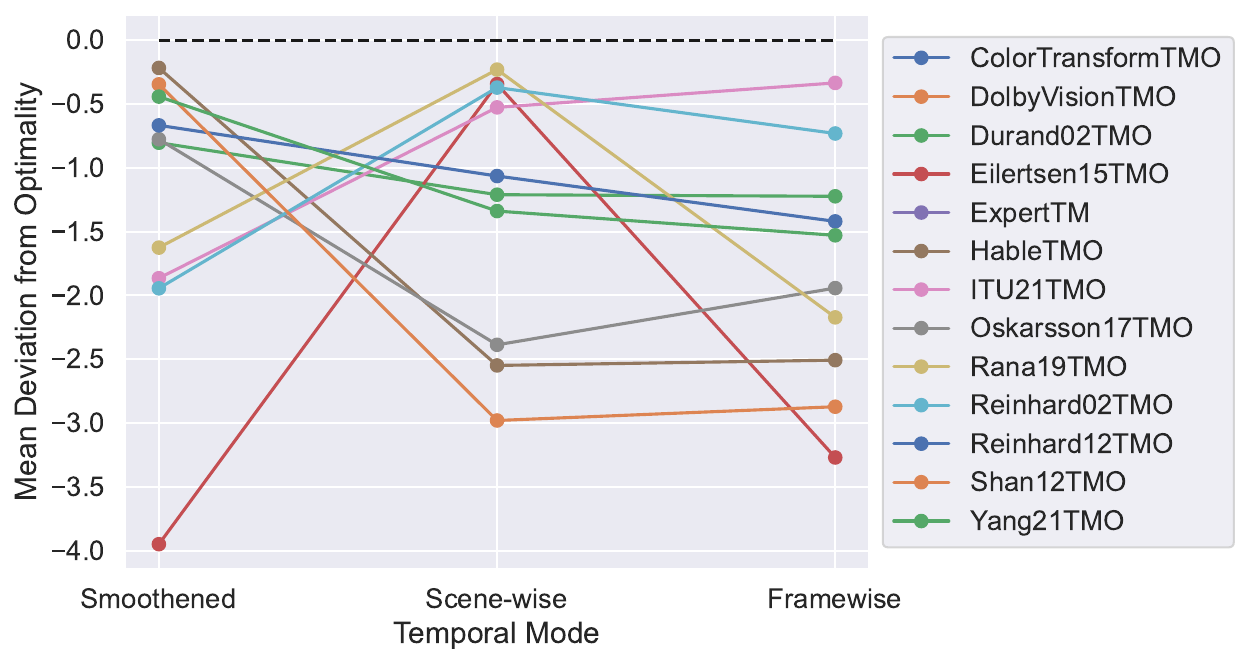}
    }
    \caption{Analyzing the impact of spatial parameters and temporal modes on quality.}
    \label{fig:spat_temp_mdo}
\end{figure*}

To further illustrate the impact of spatial and temporal parameters, we conducted a similar MDO analysis for the temporal modes and the desaturation parameter used by four TMOs. The results of the MDO analysis are shown in Fig. \ref{fig:spat_temp_mdo}, from which it may be seen that the optimal spatial parameter or temporal mode choice depends on the TMO. When analyzing the impact of desaturation, the Rana19 and Yang21 TMOs benefited from a small amount of color correction (desaturation of 0.25), while Reinhard02 and Hable did not. Similarly, the choice of the best temporal mode is either smoothing or scene-wise tone mapping for most TMOs. The only TMO that was best applied framewise was ITU21, which did not have any parameters in its original form.

\section{Evaluating Objective Quality Models}
\label{sec:evaluation}
To illustrate the usefulness of the LIVE-TMHDR database, we evaluated 15 TM-HDR quality prediction models from the literature. TMQI \cite{ref:tmqi}, FSITM \cite{ref:fsitm}, and FFTMI \cite{ref:fftmi} are full-reference image quality assessment (FR IQA) models, TMVQI \cite{ref:tmvqi}, FUNQUE+ \cite{ref:funque_plus}, and Cut-FUNQUE \cite{ref:cut_funque} are FR video quality assessment (FR VQA) models, and BRISQUE \cite{ref:brisque}, NIQE \cite{ref:niqe}, DIIVINE \cite{ref:diivine}, BTMQI \cite{ref:btmqi}, RcNet \cite{ref:rcnet}, HIGRADE \cite{ref:higrade}, and MSML \cite{ref:msml} are no-reference (NR) IQA models. We adapted all IQA models to videos by applying them framewise.

Of these, Cut-FUNQUE is a recent FR video quality model we developed in \cite{ref:cut_funque} that targets quality assessment of tone-mapped and compressed HDR videos. Cut-FUNQUE utilizes three key components to achieve SOTA quality prediction accuracy. The first is a perceptually uniform representation of color signals, which enables meaningful comparisons between HDR and SDR color stimuli. The second is the use of a binned weighting mechanism to isolate image regions having different visual characteristics, such as brightness, contrast, and temporal complexity. Finally, Cut-FUNQUE also uses a set of statistical similarity measures to compare HDR and SDR video frames, which overcomes the barriers of pixel-based comparisons across dynamic ranges. 

Each of the aforementioned methods generates a set of quality-aware features that must be mapped to quality predictions using trained regressors. To present each method, i.e., feature set, in the best light, we experimented with three regressors - Linear Support Vector Regressors (SVRs), Gaussian SVRs, and Random Forest Regressors wherever feasible, and we chose the most accurate resulting model. The accuracy of each feature set and regressor was evaluated using content-separated cross-validation, and the median Pearson Correlation Coefficient (PCC), Spearman's Rank Order Correlation Coefficient (SROCC), and Root Mean Square Error (RMSE) over 100 random 80-20 train-test splits of the database are reported in Table \ref{tab:comparison}. ``Content separation'' refers to the property that random train-test splits were generated while ensuring that the same source content was not present in both the training and test splits. Due to the large number of features in the Cut-FUNQUE model, only the Random Forest regressor could be trained.

\begin{table}[b]
    \centering
    \caption{Evaluation of Quality Prediction Models \\ in Terms of Median Cross-Validation Accuracy, \\ Including Regressor And Hyperparameter Tuning}
    \label{tab:comparison}
    \begin{tabular}{|c|c|c|c|c|}
        \hline
         Model & Regressor & PCC & SROCC & RMSE \\
         \hline
        Y-FUNQUE+ \cite{ref:funque_plus} & RandomForest & 0.4524 & 0.4343 & 9.4352 \\
        BTMQI \cite{ref:btmqi} & GaussianSVR & 0.4705 & 0.4663 & 9.2238 \\
        FSITM \cite{ref:fsitm} & LinearSVR & 0.4813 & 0.4626 & 8.9212 \\
        NIQE \cite{ref:niqe} & GaussianSVR & 0.4805 & 0.4746 & 9.5563 \\
        BRISQUE \cite{ref:brisque} & LinearSVR & 0.4811 & 0.4833 & 8.9869 \\
        DIIVINE \cite{ref:diivine} & GaussianSVR & 0.4794 & 0.4925 & 9.2879 \\
        TMQI \cite{ref:tmqi} & GaussianSVR & 0.5062 & 0.4956 & 8.6897 \\
        FUNQUE \cite{ref:funque} & RandomForest & 0.5082 & 0.4949 & 8.8863 \\
        TMVQI \cite{ref:tmvqi} & RandomForest & 0.5198 & 0.4969 & 8.8697 \\
        FFTMI \cite{ref:fftmi} & GaussianSVR & 0.5298 & 0.5315 & 8.8559 \\
        3C-FUNQUE+ \cite{ref:funque_plus} & RandomForest & 0.5817 & 0.5661 & 8.6568 \\
        RcNet \cite{ref:rcnet} & Random Forest & 0.5985 & 0.5824 & 8.2417 \\
        HIGRADE \cite{ref:higrade} & GaussianSVR & 0.6682 & 0.6698 & 8.2619 \\
        \textbf{MSML} \cite{ref:msml} & Linear SVR & \textbf{0.7883} & \textbf{0.7740} & 6.8090 \\
        \textbf{Cut-FUNQUE} \cite{ref:cut_funque} & Random Forest & 0.7783 & \textbf{0.7781} & \textbf{6.4187} \\
         \hline
    \end{tabular}
\end{table}

\begin{table*}[b]
    \centering
    \caption{Results of Welch's t-test Applied to Pairs of Objective Quality Models. \\ An entry of ``1'' (``0'') denotes that the model in the row achieves superior (inferior) accuracy to the model in the column, \\ while an entry of ``-'' denotes that the difference in accuracy is not statistically significant.}
    \label{tab:stat_comparison}
    \tiny
    \begin{tabular}{|c|ccccccccccccccc|}
    \hline
 & Y-FUNQUE+ & BTMQI & FSITM & NIQE & BRISQUE & DIIVINE & TMQI & FUNQUE & TMVQI & FFTMI & 3C-FUNQUE+ & RcNet & HIGRADE & MSML & Cut-FUNQUE\\
    \hline
    Y-FUNQUE+ & -  & -  & -  & -  & 0 & 0 & 0 & 0 & 0 & 0 & 0 & 0 & 0 & 0 & 0\\
    BTMQI & -  & -  & -  & -  & 0 & 0 & 0 & 0 & 0 & 0 & 0 & 0 & 0 & 0 & 0\\
    FSITM & -  & -  & -  & -  & -  & -  & 0 & 0 & 0 & 0 & 0 & 0 & 0 & 0 & 0\\
    NIQE & -  & -  & -  & -  & -  & -  & 0 & 0 & 0 & 0 & 0 & 0 & 0 & 0 & 0\\
    BRISQUE & 1 & 1 & -  & -  & -  & -  & -  & -  & -  & 0 & -  & 0 & 0 & 0 & 0\\
    DIIVINE & 1 & 1 & -  & -  & -  & -  & -  & -  & -  & 0 & -  & 0 & 0 & 0 & 0\\
    TMQI & 1 & 1 & 1 & 1 & -  & -  & -  & -  & -  & -  & -  & 0 & 0 & 0 & 0\\
    FUNQUE & 1 & 1 & 1 & 1 & -  & -  & -  & -  & -  & 0 & -  & 0 & 0 & 0 & 0\\
    TMVQI & 1 & 1 & 1 & 1 & -  & -  & -  & -  & -  & 0 & -  & 0 & 0 & 0 & 0\\
    FFTMI & 1 & 1 & 1 & 1 & 1 & 1 & -  & 1 & 1 & -  & -  & 0 & 0 & 0 & 0\\
    3C-FUNQUE+ & 1 & 1 & 1 & 1 & -  & -  & -  & -  & -  & -  & -  & -  & 0 & 0 & 0\\
    RcNet & 1 & 1 & 1 & 1 & 1 & 1 & 1 & 1 & 1 & 1 & -  & -  & 0 & 0 & 0\\
    HIGRADE & 1 & 1 & 1 & 1 & 1 & 1 & 1 & 1 & 1 & 1 & 1 & 1 & -  & 0 & 0\\
    MSML & 1 & 1 & 1 & 1 & 1 & 1 & 1 & 1 & 1 & 1 & 1 & 1 & 1 & -  & - \\
    Cut-FUNQUE & 1 & 1 & 1 & 1 & 1 & 1 & 1 & 1 & 1 & 1 & 1 & 1 & 1 & -  & - \\
    \hline
    \end{tabular}
\end{table*}
From Table \ref{tab:comparison}, it may be seen that Cut-FUNQUE significantly outperformed, by over 15\%, nearly all of the compared quality models on the LIVE-TMHDR database in terms of prediction accuracy. The only existing quality model rivaling Cut-FUNQUE is the deep MSML model that computes a set of 9216 features from a pre-trained ResNet-50 model. By contrast, Cut-FUNQUE relies on an efficient Haar wavelet transform and computation sharing between features. As a result, Cut-FUNQUE offers a 20\(\times\) speedup over MSML in terms of the number of floating-point operations (FLOPs) \cite{ref:cut_funque}.

To lend further weight to the comparison of algorithms, we conducted one-sided Welch's t-tests to evaluate the statistical significance of the observed differences in prediction accuracies. A one-sided Welch's t-test is preferred over a traditional Student's t-test, since a Welch's t-test accounts for unequal population variances \cite{ref:welch}. Table \ref{tab:stat_comparison} presents the results of pairwise statistical significance comparisons. An entry of ``1'' (``0'') denotes that the quality model in the row achieved statistically significantly superior (inferior) accuracy compared to the quality model in the column. An entry of ``-'' denotes that the differences are not statistically significant. From this table, it may be seen that MSML and Cut-FUNQUE outperformed all other quality models, and that the difference between the two top performers was not statistically significant.

\section{Conclusion}
In this work, we have studied the problem of subjective quality assessment of compressed tone-mapped HDR videos. To this end, we developed the first large-scale publicly available database of compressed tone-mapped videos, called the LIVE-TMHDR database. This resource consists of 40 source contents, including both user and professionally-generated content, tone mapped using 12 algorithms and a human expert, and subjected to three compression levels. In total, the database contains 15,000 test videos. The sequence of operations used to generate the test contents simulates a real-world server-side pipeline that aims to deliver HDR videos to consumers using SDR displays. 

We then conducted a large-scale subjective study using the Amazon Mechanical Turk platform to obtain crowd-sourced ground-truth annotations. Overall, we obtained over 750,000 ratings at an average of over 50 per video. Using statistical modeling techiques such as SUREAL, we used the subjective data to analyze the effect of various TMOs and the choices of their parameters on visual quality.

Finally, we utilized the subjectively-annotated database to compare the accuracies of fifteen full-reference and no-reference video quality prediction models. Through this analysis, we found that the deep-learning model MSML and the Cut-FUNQUE quality model both achieved SOTA accuracy on the proposed benchmark. We believe that this large-scale database will spur further research into HDR content delivery, including the development of novel tone-mapping algorithms and quality prediction models.

\label{sec:conclusion}

\bibliographystyle{IEEEtran}
\bibliography{refs}

\begin{thebibliography}{10}
\providecommand{\url}[1]{#1}
\csname url@samestyle\endcsname
\providecommand{\newblock}{\relax}
\providecommand{\bibinfo}[2]{#2}
\providecommand{\BIBentrySTDinterwordspacing}{\spaceskip=0pt\relax}
\providecommand{\BIBentryALTinterwordstretchfactor}{4}
\providecommand{\BIBentryALTinterwordspacing}{\spaceskip=\fontdimen2\font plus
\BIBentryALTinterwordstretchfactor\fontdimen3\font minus \fontdimen4\font\relax}
\providecommand{\BIBforeignlanguage}[2]{{%
\expandafter\ifx\csname l@#1\endcsname\relax
\typeout{** WARNING: IEEEtran.bst: No hyphenation pattern has been}%
\typeout{** loaded for the language `#1'. Using the pattern for}%
\typeout{** the default language instead.}%
\else
\language=\csname l@#1\endcsname
\fi
#2}}
\providecommand{\BIBdecl}{\relax}
\BIBdecl

\bibitem{ref:rec_709}
{ITU-R}, ``{ITU-R BT.709: Parameter values for the HDTV standards for production and international programme exchange},'' 2011.

\bibitem{ref:srgb}
IEC, ``{Multimedia systems and equipment - Colour measurement and management - Part 2-1: Colour management - Default RGB colour space - sRGB},'' 1999.

\bibitem{ref:rec_2100}
{ITU-R}, ``{ITU-R BT.2100: Image parameter values for high dynamic range television for use in production and international programme exchange},'' 2018.

\bibitem{ref:pq}
SMPTE, ``High dynamic range electro-optical transfer function of mastering reference displays,'' \emph{SMPTE Standard}, vol. 2084, no. 2014, p.~11, 2014.

\bibitem{ref:hlg}
T.~Borer and A.~Cotton, ``{A Display-Independent High Dynamic Range Television System},'' \emph{SMPTE Motion Imaging Journal}, vol. 125, no.~4, pp. 50--56, 2016.

\bibitem{ref:hdr10}
\BIBentryALTinterwordspacing
CTA. {Television technology consumer definitions}. [Online]. Available: \url{https://cdn.cta.tech/cta/media/media/membership/pdfs/videotechnology-consumer-definitions.pdf}
\BIBentrySTDinterwordspacing

\bibitem{ref:hdr10_plus}
\BIBentryALTinterwordspacing
{HDR10+ Technologies, LLC}. (2019) Hdr10+ system whitepaper. [Online]. Available: \url{https://hdr10plus.org/wp-content/uploads/2019/08/HDR10_WhitePaper.pdf}
\BIBentrySTDinterwordspacing

\bibitem{ref:hlg_satellite}
\BIBentryALTinterwordspacing
C.~Forrester, ``{SkyPerfect offers UHD-HDR by DTH}.'' [Online]. Available: \url{https://advanced-television.com/2015/11/04/skyperfect-offers-uhd-hdr-by-dth/}
\BIBentrySTDinterwordspacing

\bibitem{ref:dolbyvision}
\BIBentryALTinterwordspacing
(2016). [Online]. Available: \url{https://professional.dolby.com/siteassets/pdfs/dolby-vision-whitepaper_an-introduction-to-dolby-vision_0916.pdf}
\BIBentrySTDinterwordspacing

\bibitem{ref:hdr_tv_survey}
\BIBentryALTinterwordspacing
CNET. {Best TV for 2024: We Tested Samsung, LG, TCL, Vizio and More}. [Online]. Available: \url{https://www.cnet.com/tech/home-entertainment/best-tv/}
\BIBentrySTDinterwordspacing

\bibitem{ref:sdr_majority}
\BIBentryALTinterwordspacing
------. {How To Choose an HDR Gaming Monitor}. [Online]. Available: \url{https://www.cnet.com/tech/computing/how-to-choose-an-hdr-gaming-monitor/}
\BIBentrySTDinterwordspacing

\bibitem{ref:hdrmax_spl}
J.~P. Ebenezer, Z.~Shang, Y.~Wu, H.~Wei, S.~Sethuraman, and A.~C. Bovik, ``Making video quality assessment models robust to bit depth,'' \emph{IEEE Signal Processing Letters}, vol.~30, pp. 488--492, 2023.

\bibitem{ref:cadik}
M.~Čadík, M.~Wimmer, L.~Neumann, and A.~Artusi, ``{Evaluation of HDR tone mapping methods using essential perceptual attributes},'' \emph{Computers \& Graphics}, vol.~32, no.~3, pp. 330--349, 2008.

\bibitem{ref:gma_review}
J.~Morovic and M.~R. Luo, ``{The fundamentals of gamut mapping: A survey},'' \emph{Journal of Imaging Science and Technology}, vol.~45, no.~3, pp. 283--290, 2001.

\bibitem{ref:compression_effects}
T.~Wolff, H.-H. Ho, J.~M. Foley, and S.~K. Mitra, ``H.264 coding artifacts and their relation to perceived annoyance,'' in \emph{2006 14th European Signal Processing Conference}, 2006, pp. 1--5.

\bibitem{ref:drago}
F.~Drago, W.~L. Martens, K.~Myszkowski, and H.-P. Seidel, ``{Perceptual evaluation of tone mapping operators},'' in \emph{ACM SIGGRAPH 2003 Sketches \& Applications}, 2003, pp. 1--1.

\bibitem{ref:indscal}
J.~D. Carroll and J.-J. Chang, ``{Analysis of individual differences in multidimensional scaling via an N-way generalization of “Eckart-Young” decomposition},'' \emph{Psychometrika}, vol.~35, no.~3, pp. 283--319, 1970.

\bibitem{ref:kuang}
J.~Kuang, H.~Yamaguchi, G.~M. Johnson, and M.~D. Fairchild, ``{Testing HDR Image Rendering Algorithms},'' in \emph{International Conference on Communications in Computing}, 2004.

\bibitem{ref:ledda}
P.~Ledda, A.~Chalmers, T.~Troscianko, and H.~Seetzen, ``{Evaluation of Tone Mapping Operators Using a High Dynamic Range Display},'' \emph{ACM Trans. Graph.}, vol.~24, no.~3, p. 640–648, Jul 2005.

\bibitem{ref:thurstone}
L.~L. Thurstone, ``{A law of comparative judgment.}'' \emph{Psychological review}, vol.~34, no.~4, p. 273, 1927.

\bibitem{ref:durand02}
F.~Durand and J.~Dorsey, ``{Fast Bilateral Filtering for the Display of High-Dynamic-Range Images},'' in \emph{Proceedings of the 29th Annual Conference on Computer Graphics and Interactive Techniques}, ser. SIGGRAPH '02.\hskip 1em plus 0.5em minus 0.4em\relax New York, NY, USA: Association for Computing Machinery, 2002, p. 257–266.

\bibitem{ref:reinhard02}
E.~Reinhard, M.~Stark, P.~Shirley, and J.~Ferwerda, ``{Photographic Tone Reproduction for Digital Images},'' in \emph{Proceedings of the 29th Annual Conference on Computer Graphics and Interactive Techniques}, ser. SIGGRAPH '02.\hskip 1em plus 0.5em minus 0.4em\relax Association for Computing Machinery, 2002, p. 267–276.

\bibitem{ref:icam}
\BIBentryALTinterwordspacing
G.~Johnson and M.~Fairchild, ``{Rendering HDR images},'' 2003. [Online]. Available: \url{https://scholarworks.rit.edu/other/151}
\BIBentrySTDinterwordspacing

\bibitem{ref:yoshida}
A.~Yoshida, V.~Blanz, K.~Myszkowski, and H.-P. Seidel, ``{Perceptual evaluation of tone mapping operators with real-world scenes},'' in \emph{Human Vision and Electronic Imaging X}, vol. 5666, International Society for Optics and Photonics.\hskip 1em plus 0.5em minus 0.4em\relax SPIE, 2005, pp. 192 -- 203.

\bibitem{ref:krasula_tm_db}
L.~Krasula, M.~Narwaria, K.~Fliegel, and P.~Le~Callet, ``Preference of experience in image tone-mapping: Dataset and framework for objective measures comparison,'' \emph{IEEE Journal of Selected Topics in Signal Processing}, vol.~11, no.~1, pp. 64--74, 2017.

\bibitem{ref:eilertsen_comp16}
G.~Eilertsen, J.~Unger, and R.~K. Mantiuk, ``{Evaluation of Tone Mapping Operators for HDR Video},'' in \emph{High Dynamic Range Video}.\hskip 1em plus 0.5em minus 0.4em\relax Academic Press, 2016, pp. 185--207.

\bibitem{ref:eilertsen_review17}
G.~Eilertsen, R.~K. Mantiuk, and J.~Unger, ``{A comparative review of tone-mapping algorithms for high dynamic range video},'' \emph{Computer Graphics Forum}, vol.~36, no.~2, pp. 565--592, 2017.

\bibitem{ref:tmqi}
H.~Yeganeh and Z.~Wang, ``{Objective Quality Assessment of Tone-Mapped Images},'' \emph{IEEE Transactions on Image Processing}, vol.~22, no.~2, pp. 657--667, 2013.

\bibitem{ref:espl_live}
D.~Kundu, D.~Ghadiyaram, A.~C. Bovik, and B.~L. Evans, ``{Large-Scale Crowdsourced Study for Tone-Mapped HDR Pictures},'' \emph{IEEE Transactions on Image Processing}, vol.~26, no.~10, pp. 4725--4740, 2017.

\bibitem{ref:higrade}
D.~Kundu, D.~Ghadiyaram, A.~C. Bovik, and B.~Evans, ``{No-Reference Quality Assessment of Tone-Mapped HDR Pictures},'' \emph{IEEE Transactions on Image Processing}, vol.~26, no.~6, pp. 2957--2971, 2017.

\bibitem{ref:rvtmo}
A.~Ak, A.~Goswami, W.~Hauser, P.~Le~Callet, and F.~Dufaux, ``{RV-TMO: Large-Scale Dataset for Subjective Quality Assessment of Tone Mapped Images},'' \emph{IEEE Transactions on Multimedia}, pp. 1--12, 2022.

\bibitem{ref:live_hdr}
Z.~Shang, J.~P. Ebenezer, A.~C. Bovik, Y.~Wu, H.~Wei, and S.~Sethuraman, ``{Subjective Assessment Of High Dynamic Range Videos Under Different Ambient Conditions},'' in \emph{2022 IEEE International Conference on Image Processing (ICIP)}, 2022, pp. 786--790.

\bibitem{ref:cdvl}
\BIBentryALTinterwordspacing
{The Consumer Digital Video Library}. [Online]. Available: \url{https://www.cdvl.org/}
\BIBentrySTDinterwordspacing

\bibitem{ref:sjtu_hdr}
L.~Song, Y.~Liu, X.~Yang, G.~Zhai, R.~Xie, and W.~Zhang, ``{The SJTU HDR Video Sequence Dataset},'' in \emph{Proceedings of International Conference on Quality of Multimedia Experience (QoMEX 2016)}, Lisbon, Portugal, 2016, p.~2.

\bibitem{ref:4kmedia}
\BIBentryALTinterwordspacing
{Free Ultra-HD / HDR / HLG / Dolby Vision 4K video demos}. [Online]. Available: \url{https://4kmedia.org/}
\BIBentrySTDinterwordspacing

\bibitem{ref:sandvine2023}
\BIBentryALTinterwordspacing
Global internet phenomena report 2023. [Online]. Available: \url{https://www.sandvine.com/global-internet-phenomena-report-2023}
\BIBentrySTDinterwordspacing

\bibitem{ref:vqeg_siti}
\BIBentryALTinterwordspacing
{SI/TI Tools}. [Online]. Available: \url{https://vqeg.org/VQEGSharedFiles/MeetingFiles/2021_12_UniversityOfSurry_online/VQEG_NORM_2021_204_SI-TI.pdf}
\BIBentrySTDinterwordspacing

\bibitem{ref:cf}
D.~Hasler and S.~Suesstrunk, ``Measuring colourfulness in natural images,'' \emph{Proceedings of SPIE - The International Society for Optical Engineering}, vol. 5007, pp. 87--95, 06 2003.

\bibitem{ref:ugc_vqa}
Z.~Tu, Y.~Wang, N.~Birkbeck, B.~Adsumilli, and A.~C. Bovik, ``{UGC-VQA: Benchmarking Blind Video Quality Assessment for User Generated Content},'' \emph{IEEE Transactions on Image Processing}, vol.~30, pp. 4449--4464, 2021.

\bibitem{ref:mindist_gma}
F.~Ebner and M.~D. Fairchild, ``{Gamut mapping from below: Finding minimum perceptual distances for colors outside the gamut volume},'' \emph{Color Research \& Application}, vol.~22, no.~6, pp. 402--413, 1997.

\bibitem{ref:katoh_clipping}
N.~Katoh and M.~Ito, ``{Gamut mapping for computer generated images (II)},'' in \emph{Color and Imaging Conference}, vol. 1996, no.~1.\hskip 1em plus 0.5em minus 0.4em\relax Society for Imaging Science and Technology, 1996, pp. 126--128.

\bibitem{ref:morovic_gma}
J.~Morovic and M.~R. Luo, ``{Gamut mapping algorithms based on psychophysical experiment},'' in \emph{Color and Imaging Conference}, vol. 1997, no.~1.\hskip 1em plus 0.5em minus 0.4em\relax Society for Imaging Science and Technology, 1997, pp. 44--49.

\bibitem{ref:sigmoid_gma}
G.~J. Braun and M.~D. Fairchild, ``{General-purpose gamut-mapping algorithms: Evaluation of contrast-preserving rescaling functions for color gamut mapping},'' \emph{Journal of Imaging Science and Technology}, vol.~44, no.~4, pp. 343--350, 2000.

\bibitem{ref:itu_gma}
{ITU-R}, ``{ITU-R BT.2407: Colour gamut conversion from Recommendation ITU-R BT.2020 to Recommendation ITU-R BT.709},'' 2017.

\bibitem{ref:hable}
\BIBentryALTinterwordspacing
J.~Hable. {Uncharted 2: HDR Lighting}. [Online]. Available: \url{https://www.gdcvault.com/play/1012351/Uncharted-2-HDR}
\BIBentrySTDinterwordspacing

\bibitem{ref:ffmpeg}
\BIBentryALTinterwordspacing
{FFmpeg}. [Online]. Available: \url{https://ffmpeg.org/}
\BIBentrySTDinterwordspacing

\bibitem{ref:shan12}
Q.~Shan, T.~DeRose, and J.~Anderson, ``{Tone Mapping High Dynamic Range Videos Using Wavelets},'' \emph{Pixar Technical Memo}, 2012.

\bibitem{ref:swt}
G.~P. Nason and B.~W. Silverman, \emph{{The Stationary Wavelet Transform and some Statistical Applications}}.\hskip 1em plus 0.5em minus 0.4em\relax Springer New York, 1995, pp. 281--299.

\bibitem{ref:reinhard12}
E.~Reinhard, T.~Pouli, T.~Kunkel, B.~Long, A.~Ballestad, and G.~Damberg, ``{Calibrated Image Appearance Reproduction},'' \emph{ACM Trans. Graph.}, vol.~31, no.~6, Nov 2012.

\bibitem{ref:eilertsen15}
G.~Eilertsen, R.~K. Mantiuk, and J.~Unger, ``{Real-Time Noise-Aware Tone Mapping},'' \emph{ACM Trans. Graph.}, vol.~34, no.~6, Nov 2015.

\bibitem{ref:oskarsson17}
M.~Oskarsson, ``{Temporally Consistent Tone Mapping of Images and Video Using Optimal K-means Clustering},'' \emph{Journal of Mathematical Imaging and Vision}, vol.~57, no.~2, pp. 225--238, Feb 2017.

\bibitem{ref:rana19}
A.~Rana, P.~Singh, G.~Valenzise, F.~Dufaux, N.~Komodakis, and A.~Smolic, ``{Deep Tone Mapping Operator for High Dynamic Range Images},'' \emph{IEEE Transactions on Image Processing}, vol.~29, pp. 1285--1298, 2020.

\bibitem{ref:yang21}
J.~Yang, Z.~Liu, M.~Lin, S.~Yanushkevich, and O.~Yadid-Pecht, ``{Deep Reformulated Laplacian Tone Mapping},'' 2021.

\bibitem{ref:itu21}
ITU-R, ``{ITU-R BT.2446: Methods for conversion of high dynamic range content to standard dynamic range content and vice-versa},'' 2021.

\bibitem{ref:boitard12}
R.~Boitard, K.~Bouatouch, R.~Cozot, D.~Thoreau, and A.~Gruson, ``{Temporal coherency for video tone mapping},'' in \emph{Applications of Digital Image Processing XXXV}, A.~G. Tescher, Ed., vol. 8499, International Society for Optics and Photonics.\hskip 1em plus 0.5em minus 0.4em\relax SPIE, 2012, p. 84990D.

\bibitem{ref:vqeghd3}
(2010) {Report on the validation of video quality models for high definition video content}. Video Quality Experts Group.

\bibitem{ref:livestream}
Z.~Shang, J.~P. Ebenezer, Y.~Wu, H.~Wei, S.~Sethuraman, and A.~C. Bovik, ``{Study of the Subjective and Objective Quality of High Motion Live Streaming Videos},'' \emph{IEEE Transactions on Image Processing}, vol.~31, pp. 1027--1041, 2022.

\bibitem{ref:x264}
\BIBentryALTinterwordspacing
{VideoLAN}, ``x264.'' [Online]. Available: \url{https://code.videolan.org/videolan/x264.git}
\BIBentrySTDinterwordspacing

\bibitem{ref:can_i_use_hevc}
\BIBentryALTinterwordspacing
{Can I use - HEVC/H.265 video format}. [Online]. Available: \url{https://caniuse.com/?search=H.265}
\BIBentrySTDinterwordspacing

\bibitem{ref:can_i_use_avc}
\BIBentryALTinterwordspacing
{Can I use - MPEG-4/H.264 video format}. [Online]. Available: \url{https://caniuse.com/?search=H.264}
\BIBentrySTDinterwordspacing

\bibitem{ref:masking1}
P.~Teo and D.~Heeger, ``Perceptual image distortion,'' in \emph{Proceedings of 1st International Conference on Image Processing}, vol.~2, 1994, pp. 982--986 vol.2.

\bibitem{ref:masking2}
\BIBentryALTinterwordspacing
J.~M. Foley and G.~M. Boynton, ``{New model of human luminance pattern vision mechanisms: analysis of the effects of pattern orientation, spatial phase and temporal frequency},'' in \emph{Computational Vision Based on Neurobiology}, T.~B. Lawton, Ed., vol. 2054, International Society for Optics and Photonics.\hskip 1em plus 0.5em minus 0.4em\relax SPIE, 1994, pp. 32 -- 42. [Online]. Available: \url{https://doi.org/10.1117/12.171150}
\BIBentrySTDinterwordspacing

\bibitem{ref:amt}
\BIBentryALTinterwordspacing
{Amazon Mechanical Turk}. [Online]. Available: \url{https://www.mturk.com/}
\BIBentrySTDinterwordspacing

\bibitem{ref:live_wild}
D.~Ghadiyaram and A.~C. Bovik, ``{Massive Online Crowdsourced Study of Subjective and Objective Picture Quality},'' \emph{IEEE Transactions on Image Processing}, vol.~25, no.~1, pp. 372--387, 2016.

\bibitem{ref:kadid_10k}
H.~Lin, V.~Hosu, and D.~Saupe, ``{KADID-10k: A Large-scale Artificially Distorted IQA Database},'' in \emph{2019 Eleventh International Conference on Quality of Multimedia Experience (QoMEX)}, 2019, pp. 1--3.

\bibitem{ref:live_vqc}
Z.~Sinno and A.~C. Bovik, ``{Large-Scale Study of Perceptual Video Quality},'' \emph{IEEE Transactions on Image Processing}, vol.~28, no.~2, pp. 612--627, 2019.

\bibitem{ref:patch_vq}
\BIBentryALTinterwordspacing
Z.~Ying, M.~Mandal, D.~Ghadiyaram, and A.~C. Bovik, ``{Patch-VQ: 'Patching Up' the Video Quality Problem},'' \emph{CoRR}, vol. abs/2011.13544, 2020. [Online]. Available: \url{https://arxiv.org/abs/2011.13544}
\BIBentrySTDinterwordspacing

\bibitem{ref:konvid_150k}
F.~G\"otz-Hahn, V.~Hosu, H.~Lin, and D.~Saupe, ``{KonVid-150k: A Dataset for No-Reference Video Quality Assessment of Videos in-the-Wild},'' in \emph{IEEE Access 9}.\hskip 1em plus 0.5em minus 0.4em\relax IEEE, 2021, pp. 72\,139--72\,160.

\bibitem{ref:sureal}
Z.~Li, C.~G. Bampis, L.~Janowski, and I.~Katsavounidis, ``{A simple model for subject behavior in subjective experiments},'' \emph{Electronic Imaging}, vol. 2020, no.~11, pp. 131--1, 2020.

\bibitem{ref:fsitm}
H.~{Ziaei Nafchi}, A.~{Shahkolaei}, R.~{Farrahi Moghaddam}, and M.~{Cheriet}, ``{FSITM: A Feature Similarity Index For Tone-Mapped Images},'' \emph{IEEE Signal Processing Letters}, vol.~22, no.~8, pp. 1026--1029, 2015.

\bibitem{ref:fftmi}
L.~Krasula, K.~Fliegel, and P.~Le~Callet, ``{FFTMI: Features Fusion for Natural Tone-Mapped Images Quality Evaluation},'' \emph{IEEE Transactions on Multimedia}, vol.~22, no.~8, pp. 2038--2047, 2020.

\bibitem{ref:tmvqi}
H.~Yeganeh, S.~Wang, K.~Zeng, M.~Eisapour, and Z.~Wang, ``{Objective quality assessment of tone-mapped videos},'' in \emph{2016 IEEE International Conference on Image Processing (ICIP)}, 2016, pp. 899--903.

\bibitem{ref:funque_plus}
A.~K. Venkataramanan, C.~Stejerean, I.~Katsavounidis, and A.~C. Bovik, ``{One Transform To Compute Them All: Efficient Fusion-Based Full-Reference Video Quality Assessment},'' 2023.

\bibitem{ref:cut_funque}
A.~K. Venkataramanan, C.~Stejerean, I.~Katsavounidis, H.~Tmar, and A.~C. Bovik, ``{Cut-FUNQUE}: Objective quality assessment of compressed and tone mapped high dynamic range videos,'' \emph{Manuscript Under Preparation}, vol.~1, 2024.

\bibitem{ref:brisque}
A.~Mittal, A.~K. Moorthy, and A.~C. Bovik, ``{No-Reference Image Quality Assessment in the Spatial Domain},'' \emph{IEEE Transactions on Image Processing}, vol.~21, no.~12, pp. 4695--4708, 2012.

\bibitem{ref:niqe}
A.~{Mittal}, R.~{Soundararajan}, and A.~C. {Bovik}, ``{Making a ``Completely Blind” Image Quality Analyzer},'' \emph{IEEE Signal Processing Letters}, vol.~20, no.~3, pp. 209--212, 2013.

\bibitem{ref:diivine}
A.~K. Moorthy and A.~C. Bovik, ``Blind image quality assessment: From natural scene statistics to perceptual quality,'' \emph{IEEE Transactions on Image Processing}, vol.~20, no.~12, pp. 3350--3364, 2011.

\bibitem{ref:btmqi}
K.~Gu, S.~Wang, G.~Zhai, S.~Ma, X.~Yang, W.~Lin, W.~Zhang, and W.~Gao, ``{Blind Quality Assessment of Tone-Mapped Images Via Analysis of Information, Naturalness, and Structure},'' \emph{IEEE Transactions on Multimedia}, vol.~18, no.~3, pp. 432--443, 2016.

\bibitem{ref:rcnet}
C.~S. Ravuri, R.~Sureddi, S.~V.~R. Dendi, S.~Raman, and S.~S. Channappayya, ``{Deep no-reference tone mapped image quality assessment},'' in \emph{2019 53rd Asilomar Conference on Signals, Systems, and Computers}.\hskip 1em plus 0.5em minus 0.4em\relax IEEE, 2019, pp. 1906--1910.

\bibitem{ref:msml}
Q.~He, D.~Li, T.~Jiang, and M.~Jiang, ``{Quality Assessment for Tone-Mapped HDR Images Using Multi-Scale and Multi-Layer Information},'' in \emph{2018 IEEE International Conference on Multimedia \& Expo Workshops (ICMEW)}, 2018, pp. 1--6.

\bibitem{ref:funque}
A.~K. Venkataramanan, C.~Stejerean, and A.~C. Bovik, ``{FUNQUE: Fusion of Unified Quality Evaluators},'' in \emph{2022 IEEE International Conference on Image Processing (ICIP)}, 2022, pp. 2147--2151.

\bibitem{ref:welch}
M.~Delacre, D.~Lakens, and C.~Leys, ``Why psychologists should by default use welch's t-test instead of student's t-test,'' \emph{International Review of Social Psychology}, vol.~30, no.~1, pp. 92--101, 2017.

\end{thebibliography}
\end{document}